\documentclass[twocolumn]{aastex61}

\usepackage{colortbl}
\usepackage{soul}
\usepackage{amsmath}
\usepackage[bottom]{footmisc}

\definecolor{cnn1d}{rgb}{0.00784313725490196,0.6196078431372549,0.45098039215686275}
\definecolor{cnn2d}{rgb}{0.00392156862745098,0.45098039215686275,0.6980392156862745}
\definecolor{msigpure}{rgb}{0.8,0.47058823529411764,0.7372549019607844}
\definecolor{msigcontam}{rgb}{0.8352941176470589,0.3686274509803922,0.0}
\definecolor{sdm1d}{rgb}{0.9254901960784314,0.8823529411764706,0.2}
\definecolor{sdm2d}{rgb}{0.984313725490196,0.6862745098039216,0.8941176470588236}

\submitjournal{ApJ}

\shortauthors{}

\begin{document}

\title{A Robust and Efficient Deep Learning Method for Dynamical Mass Measurements of Galaxy Clusters}

\begin{abstract}
We demonstrate the ability of convolutional neural networks (CNNs) to mitigate systematics in the virial scaling relation and produce dynamical mass estimates of galaxy clusters with remarkably low bias and scatter. We present two models, CNN$_\mathrm{1D}$ and CNN$_\mathrm{2D}$, which leverage this deep learning tool to infer cluster masses from distributions of member galaxy dynamics. Our first model, CNN$_\text{1D}$, infers cluster mass directly from the distribution of member galaxy line-of-sight velocities. Our second model, CNN$_\text{2D}$, extends the input space of CNN$_\text{1D}$ to learn on the joint distribution of galaxy line-of-sight velocities and projected radial distances. We train each model as a regression over cluster mass using a labeled catalog of realistic mock cluster observations generated from the MultiDark simulation and UniverseMachine catalog. We then evaluate the performance of each model on an independent set of mock observations selected from the same simulated catalog. The CNN models produce cluster mass predictions with lognormal residuals of scatter as low as $0.132$ dex, greater than a factor of 2 improvement over the classical $M$-$\sigma$ power-law estimator. Furthermore, the CNN model reduces prediction scatter relative to similar machine learning approaches by up to $17\%$ while executing in drastically shorter training and evaluation times (by a factor of 30) and producing considerably more robust mass predictions (improving prediction stability under variations in galaxy sampling rate by $30\%$). 
\end{abstract}


\section{Introduction} \label{sec:intro}

Galaxy clusters are the most massive gravitationally bound structures in the universe. Clusters are complex, dark-matter-dominated systems of mass $\gtrsim 10^{14}\ h^{-1}\mathrm{M}_\odot$. Galaxy clusters dominate the high-mass tail of the halo mass function (HMF) and cluster number density is a highly sensitive probe of the growth of structure. Because of this distinction, measurements of galaxy cluster abundance as a function of mass and redshift are a major method to test cosmological models \citep[e.g.][]{Voit2005,Allen2011, Mantz2015, Planck2016}.

Utilizing cluster abundance in precision cosmology requires a large, well-defined cluster sample and robust mass measurement methods. Furthermore, modern cluster measurement techniques are expected to place a strong emphasis on efficiency and automation, as the wealth of detailed cluster data is expected to greatly increase with current and upcoming surveys such as DES, LSST, WFIRST, and \textit{Euclid} \citep{Dodelson2016}. Current methods infer cluster masses from one of several mass-dependent observables, which occur at a variety of wavelengths, including the emission of X-rays by hot intracluster gas \citep[e.g.][]{Mantz2016, Giles2017}, the scattering of CMB photons on intracluster plasma \citep[e.g.][]{SZ1972, Planck2016}, the gravitational lensing of background light \citep[e.g.][]{Applegate2014,McClintock2019}, and the properties of luminous member galaxies \citep[e.g.][]{Old2014}. Galaxy-based techniques probe clusters using multiband and spectroscopic measurements, relating mass to cluster features such as richness \citep[e.g.][]{Yee2003, Baxter2016, Old2014}, escape velocity profile \citep[e.g.][]{Diaferio1997, Diaferio1999, Gifford2013}, and member dynamics \citep[e.g.][]{Gerke2005,Old2014}. For an extensive review and comparison of galaxy-based techniques, see \citet{Old2014}.

Dynamical mass measurements are a broad classification of galaxy-based techniques which infer cluster mass from the line-of-sight (LOS) velocity distribution of galaxies. The classical approach for dynamical measurements is the $M$-$\sigma$ scaling relation, which connects a virialized cluster's total mass to the velocity dispersion of its galaxies via a power law \citep[e.g.][]{Evrard2008}. Dynamical measurements of this nature were famously used to infer the existence of dark matter in the Coma cluster \citep{Zwicky1933}. While historically significant, the $M$-$\sigma$ relation makes several assumptions about clusters which are unreliable in practice, including spherical symmetry, gravitational equilibrium, and perfect member selection. In reality, proper modeling of clusters requires careful consideration of systematics such as dynamical substructure \citep[e.g.][]{Saro2013,Wojtak2013, Old2018}, halo environment \citep[e.g.][]{White2010}, triaxiality \citep[e.g.][]{Skielboe2012, Saro2013, Svensmark2015}, and mergers \citep[e.g.][]{Evrard2008, Ribeiro2011}. In addition, galaxy selection effects are a primary source of scatter in dynamical mass predictions, as the member sample can be incomplete or otherwise contaminated by unbound interloper galaxies \citep{Saro2013,Old2015,Wojtak2018}. Modern applications of the $M$-$\sigma$ relation mitigate these effects using complex membership modeling and interloper removal schemes \citep[e.g.][]{Wojtak2007, Mamon2013, Farahi2016, Farahi2018, Abdullah2018}. 

Recently, a suite of machine-learning (ML) algorithms have been used to reconstruct dynamical cluster masses. This class of methods often involves training an ML model on a large data set of simulation-generated mock observations to then produce inference on unlabeled observations. \citet{Ntampaka2015, Ntampaka2016} introduced an ML method to infer mass from the full LOS velocity distribution of cluster members. This method attempts to capture higher-order features of the velocity distribution using a support distribution machine \citep[SDM; ][]{Sutherland2012} and has been shown to reduce scatter of traditional dynamical mass predictions ($M$-$\sigma$) by a factor of 2. \citet{Armitage2018} applied a variety of simple regression models on a hand-built feature set of dynamics observables to achieve similar error margins. \citet{Calderon2019} regressed mass on a list of cluster properties via several more complex ML models (XGBoost, Random Forests, and neural networks) to ultimately achieve prediction improvements comparable to previous ML approaches. \citeauthor{Calderon2019} briefly discussed the impacts of simulation assumptions on ML model fitting and produced preliminary predictions on cluster observations from SDSS.
 
In this paper, we introduce a novel deep learning methodology for measuring cluster masses from galaxy dynamics. The core of our model is a convolutional neural network (CNN), a deep learning tool which has received considerable attention for its applications in image recognition. We utilize kernel density estimators (KDEs) to create phase-space mappings of each cluster's galaxy dynamics distribution which serve as ``image" inputs to our CNNs. We train CNNs as a regression over logarithmic cluster mass using a catalog of realistic mock observations. We then use the trained CNN models to perform inference on unseen mock test data to evaluate model performance. This paper is organized into the following sections: In Section \ref{sec:dataset}, we discuss our simulation, galaxy labeling, and mock observation procedures. In Section \ref{sec:method}, we discuss the background and methodology surrounding application of our machine-learning algorithm. In Section \ref{sec:compare}, we describe details of several comparative methods which will serve as a baseline for evaluating the performance of our model. In Section \ref{sec:results}, we discuss performance metrics and evaluate the performance of our model. We summarize conclusions in Section \ref{sec:conclusion}. Lastly, we provide an appendix describing the explicit calculations of our mock observables (\hyperref[apx]{Appendix}). The code developed for this analysis has been made publicly available on Github\footnote{\href{https://github.com/McWilliamsCenter/halo_cnn}{https://github.com/McWilliamsCenter/halo\_cnn}}.

\section{Data Set}\label{sec:dataset}
In this section, we discuss the creation of our data set, namely the calculation of mock cluster observations. Clusters and galaxies are modeled as dark matter halos present in a $z=0.117$ snapshot of the MultiDark Planck 2 $N$-body simulation \citep{Klypin2016}. Simulated clusters are converted to realistic mock observables in agreement with the simulation's original cosmology. Mock cluster observations are designed to include realistic systematics which would impact dynamical mass estimates.

\subsection{Simulation and Galaxy Assembly} \label{subsec:simulation}
The mock observations were created using data from the MultiDark Planck 2 simulation \citep[MDPL2; ][]{Klypin2016}. MDPL2 is a large $N$-body dark matter simulation which evolves $3840^3$ particles from $z=120$ to $z=0$ within a box length of $1\ h^{-1}\mathrm{Gpc}$ and at a mass resolution of $1.51\times10^9\ h^{-1}\mathrm{M}_\odot$. The force resolution varies from 13 $h^{-1}\mathrm{kpc}$ at high z to 5 $h^{-1}\mathrm{kpc}$ at low z. The simulation is executed using the publicly available L-GADGET-2 code \citep{Springel2005} and uses a $\Lambda$CDM cosmology consistent with 2013 Planck data \citep{Planck2014}: $\Omega_\Lambda=0.693$, $\Omega_m=0.307$, $h=0.678$, $n=0.96$, $\sigma_8=0.8228$.

We model clusters and their member galaxies as host halos and subhalos, respectively. We utilize a halo catalog generated from MDPL2 simulation data using the ROCKSTAR halo finder \citep[MDPL2 Rockstar; ][]{Behroozi2013}. The MDPL2 Rockstar catalog identifies a hierarchy of host halos and subhalos within the MDPL2 simulation at sequential redshift snapshots throughout the simulation evolution. Clusters are painted onto host halos, inheriting properties such as mass, radius, position, and velocity. The mass definition applied for our simulated clusters is $M_\text{200c}$, calculated via spherical overdensities of 200 times the critical density of the MDPL2 simulation. Galaxies are painted onto subhalos through the galaxy assignment procedure UniverseMachine \citep{Behroozi2019}. By tracking the gravitational evolution of disrupted halos below the resolution limit of ROCKSTAR, UniverseMachine produces a rich and detailed catalog of simulated galaxies ideal for our data set. UniverseMachine determines stellar formation rates and masses for each galaxy, which are consistent with observational constraints. UniverseMachine galaxies inherit position and velocity from their associated subhalos.

We conduct this analysis on a publicly available $z=0.117$ snapshot of the MDPL2 simulation\footnote{\href{https://www.cosmosim.org/}{https://www.cosmosim.org/}}. The MDPL2 Rockstar and UniverseMachine catalogs provide mass, comoving position, and proper velocity information for host halos and subhalos. Host halos included in our sample are constrained to $M_{200c} \geq 10^{13.5}\ h^{-1}\mathrm{M}_\odot$. Galaxy subhalos in our sample are restricted to a stellar mass limit of $M_\mathrm{stellar}\geq 10^{9.5}\ h^{-1}\mathrm{M}_{\odot}$.

\subsection{Contaminated Mock Observations}\label{subsec:contam_mocks}

The mock observations are designed to model physical and selection effects inherent in real cluster measurements. The physical effects (cluster mergers, triaxiality), are encoded in the distributions of cluster members and surrounding material. The selection effects (interlopers) arise from nonmember galaxies positioned along the LOS and with similar perceived LOS velocities to the host cluster. To account for these effects, the mock observations select samples of member galaxies by taking large, fixed-size cylindrical cuts positioned at the cluster center and oriented along the LOS axis. This cut allows information regarding interlopers and cluster shape to contaminate the sample. We will refer to the realistic mock observations as the \textit{contaminated} catalog. A previous version of the mock observation procedure used in this paper is described in \citet{Ntampaka2016}.

In creating this set of mock cluster observations, we make the following assumptions: \textbf{(1)} All subhalos tracked by UniverseMachine above $M_\mathrm{stellar} \geq 10^{9.5}\ h^{-1} \mathrm{M}_{\odot}$ are assumed to represent a galaxy, with the galaxy inheriting its subhalo's position and velocity. \textbf{(2)} Host halos with mass $M_{200c} \geq 10^{13.5}\ h^{-1}\mathrm{M}_\odot$ are considered to be cluster candidates. \textbf{(3)} The cluster center is assumed to be known and consistent with the host halo's position and velocity. \textbf{(4)} Each cluster observation considers a unique observer assumed to lie at $z=0$ along the chosen LOS. Obstructions, lensing, and other observational artifacts are not accounted for.

Before observational cuts are calculated, the simulation snapshot box is padded on each side to account for periodic boundary conditions. At each box face and edge, a slice of galaxy data is duplicated from across the periodic boundary. The padding width is calculated from simulation data and overestimated so as to not exclude any galaxies which might be captured in a cluster's cylinder cut. This analysis used a padding width of $112\ h^{-1}\mathrm{Mpc}$. This creates a final padded cube of side length $1.224\ h^{-1}\mathrm{Gpc}$.

For a given LOS axis (\S\ref{subsec:traintest}), we determine cluster membership by first calculating the position and velocity observables for each cluster-galaxy pair. We calculate $x_\text{proj}$, $y_\text{proj}$, and $v_\text{los}$ for all galaxies around a cluster center, where $x_\text{proj}$ and $y_\text{proj}$ are projected plane-of-sky $x'$ and $y'$-positions and $v_\text{los}$ is the net LOS velocity. The net velocity, $v_\text{los}$, is given by the sum of the object's relative peculiar velocity and Hubble flow along the LOS. The quantities $x_\text{proj}$, $y_\text{proj}$, and $v_\text{los}$ are expressed as relative values to the cluster candidate's center. We also calculate the projected plane-of-sky radial distance $R_\text{proj}$, defined as the Euclidean distance to the cluster center. For a full description of the calculation of these mock observables, see the \hyperref[apx]{Appendix}.

\begin{figure}
    \centering
	\includegraphics[width=0.47\textwidth]{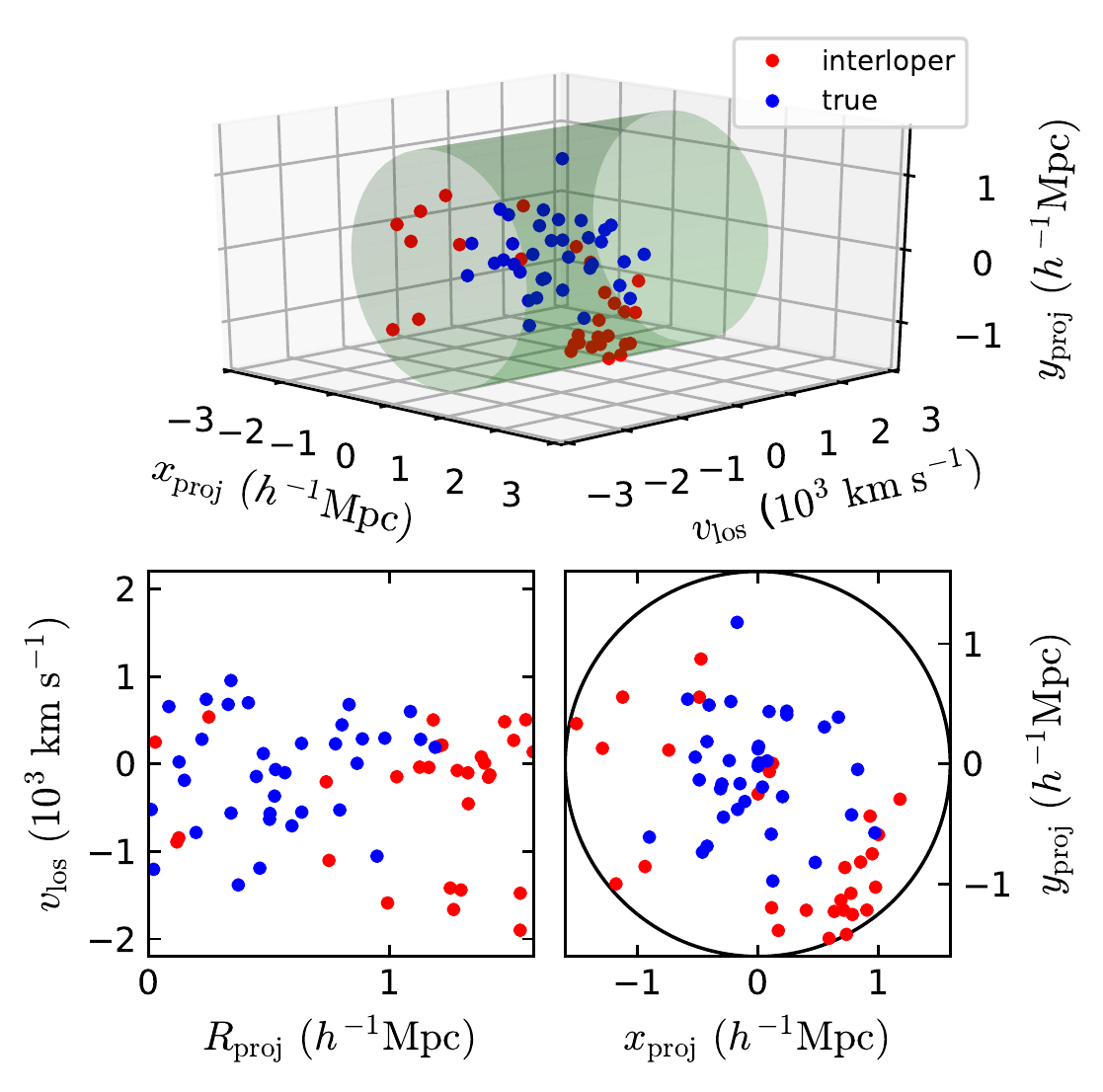}
	\caption{An example contaminated cluster member distribution showing both true members (blue) and interlopers (red), with a total $\log_{10}[M_\text{200c}\ (h^{-1}\mathrm{M}_\odot)] = 14.27$. True members correspond to galaxies that fall within the cluster's MDPL2 Rockstar FOF group. It is important to note that in our model (and in reality), we cannot distinguish between true members and interlopers. \textbf{Top}: Cluster members extracted from a cylinder cut in the mock light cone. \textbf{Bottom left}: Traditional $v_\text{los}$ vs. $R_\text{proj}$ showing cluster member distribution in projected phase space. \textbf{Bottom right}: Projected plane-of-sky perspective.}
	\label{fig:ex_cluster}
\end{figure}

The cylindrical cuts are characterized by three fixed parameters, $R_\text{aperture}$, $v_\text{cut}$, and $N_\text{min}$, which correspond to the cylinder's radial aperture in the $x_\text{proj}$-$y_\text{proj}$ plane, the half-length along the $v_\text{los}$-axis, and the minimum cluster richness, respectively. Galaxy subhalos which fall between the bounds $R_\text{proj} \leq R_\text{aperture}$ and $|v_\text{los}| \leq v_\text{cut}$ are included in the mock observation of the host cluster, whether or not they are truly gravitationally bound to the system. Following the cylindrical cut, cluster candidates that have less than $N_\text{min}$ galaxy subhalos are discarded from our sample. In this analysis, the cylinder parameters are chosen to be $R_\text{aperture}=1.6\ h^{-1}\mathrm{Mpc}$ and $v_\text{cut} = 2200\ \mathrm{km}\ \mathrm{s}^{-1}$, corresponding to the typical radius and $2\sigma_v$ of a $10^{15}\ h^{-1}\mathrm{M}_\odot$ massive halo. We use a richness cut of $N_\text{min} = 10$. The cylindrical cut procedure is symmetric for azimuthal rotations about the LOS axis. This symmetry is taken into account when augmenting training data in Section \ref{subsec:training}. An example contaminated mock observation is shown in Figure \ref{fig:ex_cluster}. 

\begin{figure}
    \centering
	\includegraphics[width=0.47\textwidth]{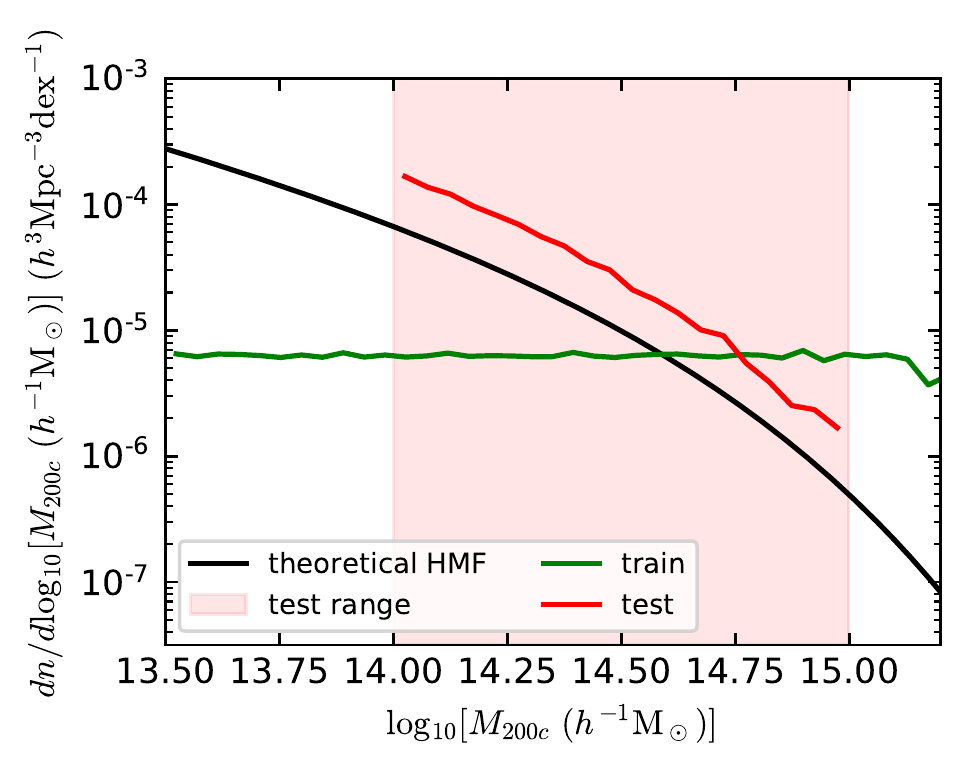}
    \caption{Mock cluster mass function for training and test samples in the contaminated catalog relative to the theoretical HMF for MDPL2 cosmology.  The test sample cluster mass function is equivalent to three times the theoretical HMF, for the three orthogonal LOS perspectives taken of each cluster. The training sample has a flat cluster mass function, to eliminate selection bias during training. Note, to create the flat mass function training set, clusters are downsampled at low masses and upsampled at high masses.}
    \label{fig:hmf}
\end{figure}

\subsection{Train/Test Split}\label{subsec:traintest}
We build a training set of mock cluster observations with a flat mock cluster mass function across all masses so as not to introduce a bias in mass predictions via an imposed prior on cluster abundance. Due to a scarcity of simulated halos above $M_\text{200c} \geq 10^{14.6}\ h^{-1}\mathrm{M}_\odot$ (Figure \ref{fig:hmf}), we create an evenly distributed training set by upsampling clusters at high masses and downsampling clusters at low masses. We execute our sampling procedure by generating new mock cluster projections from various LOS. First, we choose a number density of clusters which provides a sufficient number of cluster examples to effectively train our model without overfitting. Here, we choose a flat training cluster number density of $10^{-5.2}\ h^3\mathrm{Mpc}^{-3}\mathrm{dex}^{-1}$. Next, each cluster in our catalog is evaluated at three orthogonal LOS projections. Then, clusters in an abundant mass region are downsampled to our chosen cluster number density. Clusters in scarce mass regions are upsampled by taking additional LOS projections. Any additional LOS projections aside from the initial three are distributed with roughly even spacing on the unit sphere, according to a Fibonacci Lattice \citep{Gonzalez2010}. The average number of LOS samplings per cluster for the full training catalog is 2.91. The training set mock cluster mass function is shown in Figure \ref{fig:hmf}.

To evaluate our model under realistic measurement conditions, the test catalog cluster mass function is weighted to follow the theoretical halo mass function (HMF), i.e., the exact distribution of cluster masses that is present in our base simulation. The test set solely consists of three orthogonal LOS projections of each cluster. The testing mass range is restricted to $14\leq\log_{10}[M_{200c}\ (h^{-1}\mathrm{M}_\odot)]\leq15$ so as to avoid unreliable mean-reversion edge effects. The test set cluster mass function is shown in Figure \ref{fig:hmf}.

\subsection{Summary}
The data set generation can be summarized with the following procedure:
\begin{enumerate}
    \item MDPL2 and UniverseMachine provide position, velocity, and mass information for dark matter halos and subhalos at a chosen redshift $z=0.117$. Host halos are considered to be cluster candidates if $M_\text{200c} \geq 10^{13.5}\ h^{-1}\mathrm{M}_\odot$. Subhalos represent galaxies if they have a mass at accretion of $M_\text{acc} \geq 10^{11}\ h^{-1}\mathrm{M}_\odot$. Cluster centers are assumed to be known and consistent with the host halo's position and velocity.
    \item The simulation box is padded along each side to account for periodic boundaries. The padding width used in this analysis is overestimated at 112 $h^{-1} \mathrm{Mpc}$.
    \item Each cluster candidate's center is placed at $z=0.117$ and an observer is placed at $z=0$. The quantities $x_\text{proj}$, $y_\text{proj}$, and $v_\text{los}$ are calculated for each cluster-member pair using the procedure described in Equation (\ref{eqn:x_proj})-(\ref{eqn:v_los}).
    \item For the contaminated catalog, the mock observations consist of all galaxies within a cylinder cut of fixed radius $R_\text{aperture}$ and length $2v_\text{cut}$ centered at each cluster center in $\{x_\text{proj}, y_\text{proj}, v_\text{los}\}$-space. For the pure catalog, all galaxies within the virial radius of a given cluster are included in its mock observation. For both the pure and contaminated catalogs, all cluster candidates below a minimum richness of 10 galaxies are discarded. 
    \item Training and test sets are created from the mock observation catalogs. The training set is constructed with a flat cluster mass function in an effort to mitigate prediction bias. The test set follows the simulation's theoretical HMF. We sample the catalog to match these cluster mass function trends accordingly. Upsampling involves repeating steps 3-4 from multiple projected LOSs.
\end{enumerate}

\section{Method} \label{sec:method}
In this section, we present the deep learning methodology used to infer masses from cluster member galaxy dynamics. Our first model, CNN$_\text{1D}$, uses the distribution of galaxy line-of-sight velocities $\{v_\text{los}\}$ to infer cluster mass. This model is then extended to CNN$_\text{2D}$ by incorporating the projected plane-of-sky radius $R_\text{proj}$ as an additional input dimension. In Section \ref{subsec:KDE} we discuss how catalog data are preprocessed to serve as input to our deep learning architectures. We then describe our ML model in Section \ref{subsec:CNN} and our training/evaluation procedures in Section \ref{subsec:training}.

\subsection{Preprocessing}\label{subsec:KDE}
For each cluster, we map the distribution of member galaxies in projected phase space using a KDE. KDEs effectively smooth the distribution of discrete galaxy positions into a continuous PDF according to some prescribed length scale (bandwidth). This smoothed distribution is then sampled at regular intervals to form a pixelated mapping over the cylinder cut. PDF mappings generated with KDEs can sufficiently encapsulate features of the underlying member distribution while remaining relatively invariant to variations in the sampling rate. These mappings serve as direct input to our ML model.

\subsubsection{Kernel Density Estimation}

\begin{figure*}
    \centering 
	\includegraphics[width=\linewidth]{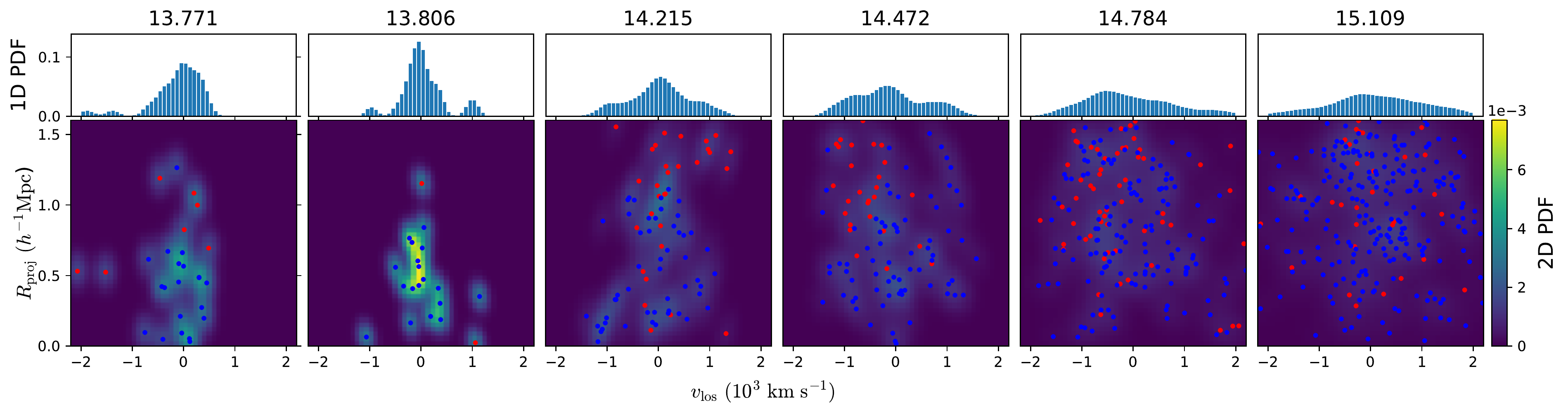}
	\caption{Six example contaminated clusters randomly selected from evenly spaced log mass bins. Each column shows the 1D and 2D normalized PDFs generated from each cluster's member distribution using a Gaussian KDE. The title of each plot gives the true $\log_{10}[M_{200c}\ (h^{-1}\mathrm{M}_\odot)]$ value assigned to each cluster. The populations of true members (blue) and interlopers (red) are superimposed on the 2D PDFs, though it is important to note that this information is not passed to the CNN models. The 1D and 2D PDFs shown here are estimated using Gaussian KDEs with a bandwidth factor of 0.25. The 1D PDFs are equivalent to the 2D PDFs marginalized over $R_\text{proj}$.}
	\label{fig:ex_kde}
\end{figure*}

Given a univariate, independent, and identically distributed sample $\{x_i\}$ of length $n$ drawn from some unknown distribution with density $f$, we can derive an expression for the estimated PDF $\hat{f}$ using a KDE,
\begin{equation}
    \hat{f}(x) = \frac{1}{n h}\sum_{i=1}^n K \Big(\frac{x-x_i}{h}\Big),
\end{equation}
where $K$ is a kernel function and $h$ is the kernel bandwidth. The kernel function is nonnegative, integrates to unity, and is often chosen to be the standard normal distribution (Gaussian KDE). The kernel bandwidth is a smoothing parameter, which we will assign to scale linearly with the sample standard deviation $h = h_0\hat{\sigma}_x$.

Product kernel estimators are used to estimate multivariate PDFs. Product kernels use the same univariate kernel in each dimension, but with a possibly different smoothing bandwidth for each dimension. Given a multivariate, independent, and identically distributed sample $\{(x_i^0,\dots, x_i^d)\}$ of length $n$ and dimension $d$ drawn from some unknown distribution with density $f$, a product kernel $\hat{f}$ can be written as
\begin{equation}
    \hat{f}(\mathbf{x}) = \frac{1}{n h_1 \cdots h_d}\sum_{i=1}^n \Bigg\{ \prod_{j=1}^d K\Big( \frac{x^j - x_i^j}{h_j}\Big)\Bigg\},
\end{equation}
where $K$ is the kernel function (like the standard normal), $\mathbf{x} = (x^0,\dots,x^d)$ is the evaluation point, and $\{h_i\}$ is the set of smoothing bandwidths. The smoothing bandwidths scale with the sample's standard deviation along their respective dimension $h_i = h_0 \sigma_i$. The bandwidth scaling factor $h_0$ is a constant coefficient applied to all smoothing bandwidths. For a comprehensive discussion of univariate and product kernels, see \citet[][chap. 6]{Scott2015}.

\subsubsection{Model Input}\label{subsubsec:input}

The CNN$_\text{1D}$ model learns on cluster $\{v_\text{los}\}$ distributions estimated using a univariate Gaussian KDE. We know from the $M$-$\sigma$ relation that the shape of the $\{v_\text{los}\}$ distribution contains information regarding the cluster mass. The set of cluster PDFs are generated at a fixed bandwidth scaling factor of $h_0=0.25$. We sample each $\{v_\text{los}\}$ PDF at 48 evenly spaced points across the cylinder cut, producing a fixed-length vector describing the distribution. Normalizing this vector to unity produces our input for the CNN$_\text{1D}$ model. Examples of the normalized $\{v_\text{los}\}$ PDF vector are shown in Figure \ref{fig:ex_kde}.

The CNN$_\text{2D}$ model uses a bivariate product kernel estimator to form a joint $\{R_\text{proj},v_\text{los}\}$ distribution. Similar to the $M$-$\sigma$ relation, the $R_\text{proj}$ distribution is descriptive of cluster mass \citep{Ntampaka2016, Armitage2018}. In addition, the joint $\{R_\text{proj},v_\text{los}\}$ shows clustering behavior of true member and interloper populations (Figure \ref{fig:ex_cluster} and \ref{fig:ex_kde}). We create a bivariate product kernel estimator for each clusters $\{R_\text{proj},v_\text{los}\}$ distribution with a fixed bandwidth scaling factor $h_0=0.25$. We sample the PDF at $48\times48$ points regularly spaced across the $\{R_\text{proj},v_\text{los}\}$ phase space. This produces a $48\times48$ array which we then normalize to unity. This array serves as input to the CNN$_\text{2D}$ model and is demonstrated in Figure \ref{fig:ex_kde}. 

\subsection{Models}\label{subsec:CNN}

The foundations of our mass estimators are CNNs. CNNs are a class of feed-forward deep neural networks (DNNs) which have garnered considerable attention recently for their applications in computer vision. CNNs have \textit{convolutional} layers that learn patterns on subsets of data. The objective of this approach is to allow convolutional layers to learn and correct for observational constraints such as interlopers and cluster mergers.

\newpage 
\subsubsection{Deep Learning}
DNNs are a group of supervised machine learning methods which encompass CNNs. DNNs have been shown to be able to learn complex, nonlinear relationships between fixed-length input and output arrays \citep{LeCun2015} and have been met with a plethora of applications in observational cosmology \citep[e.g.][]{Dielman2015,Hoyle2016,Lanusse2018,Ntampaka2018}. Within a DNN, input and output are related through a sequence of connected neuron layers. The neurons of each layer are linked to neurons of adjacent layers through a multitude of directed, weighted connections. During evaluation, each neuron produces a numerical output by taking a linear combination of values from its incoming connections and subjecting the result to a nonlinear \textit{activation function}. In the simplest case of a \textit{feed-forward} DNN, the neuron layers are evaluated in sequence, passing information from layer to layer without recurrence. Stated in tensor notation, the output $\mathbf{h}^{(l)}$ of the $l$-th layer of a feed-forward neural network can be described by the following:
\begin{equation}
    \mathbf{h}^{(l)} = f\left(\mathbf{W}^{(l)}\cdot \mathbf{h}^{(l-1)} + \mathbf{b}^{(l)}\right),
    \label{eqn:neural}
\end{equation}
where $\mathbf{W}^{(l)}$ is a matrix of connection weights, $\mathbf{b}^{(l)}$ is a vector of additive biases, and $f$ is the element-wise nonlinear activation function (e.g. sigmoid). The set $\{\mathbf{W}, \mathbf{b}\}$ constitutes the model parameters for the DNN. We consider the layers $1\leq l \leq L-1$ as part of the DNN architecture, whereas $\mathbf{h}^{(0)}$ is the input vector and $\mathbf{h}^{(L)}$ is the final, output vector. Neural layers of this form (Equation \ref{eqn:neural}) wherein every neuron is connected to every neuron of the previous layer (i.e., $\mathbf{W}^{(l)}$ is dense) are often referred to as \textit{dense} or \textit{fully connected} layers.

DNNs are trained to relate input and output by optimizing connection weights between neuron layers. During model training, we evaluate the network on a set of inputs for which the true, desired output is known. We then calculate the model's prediction error by comparing the model's output to the true values using a \textit{loss function}. We seek to minimize this prediction loss by exploring the parameter space of all connection weights using an iterative parameter optimization algorithm such as \textit{stochastic gradient descent} \citep[SGD;][]{Robbins1951}. SGD repeats its update procedure for many small, randomly selected sets of training data until the loss function stops decreasing. At this point, one might evaluate the performance of the now-optimized network on a set of independent test data.

\begin{figure*}
    \centering
    \gridline{\fig{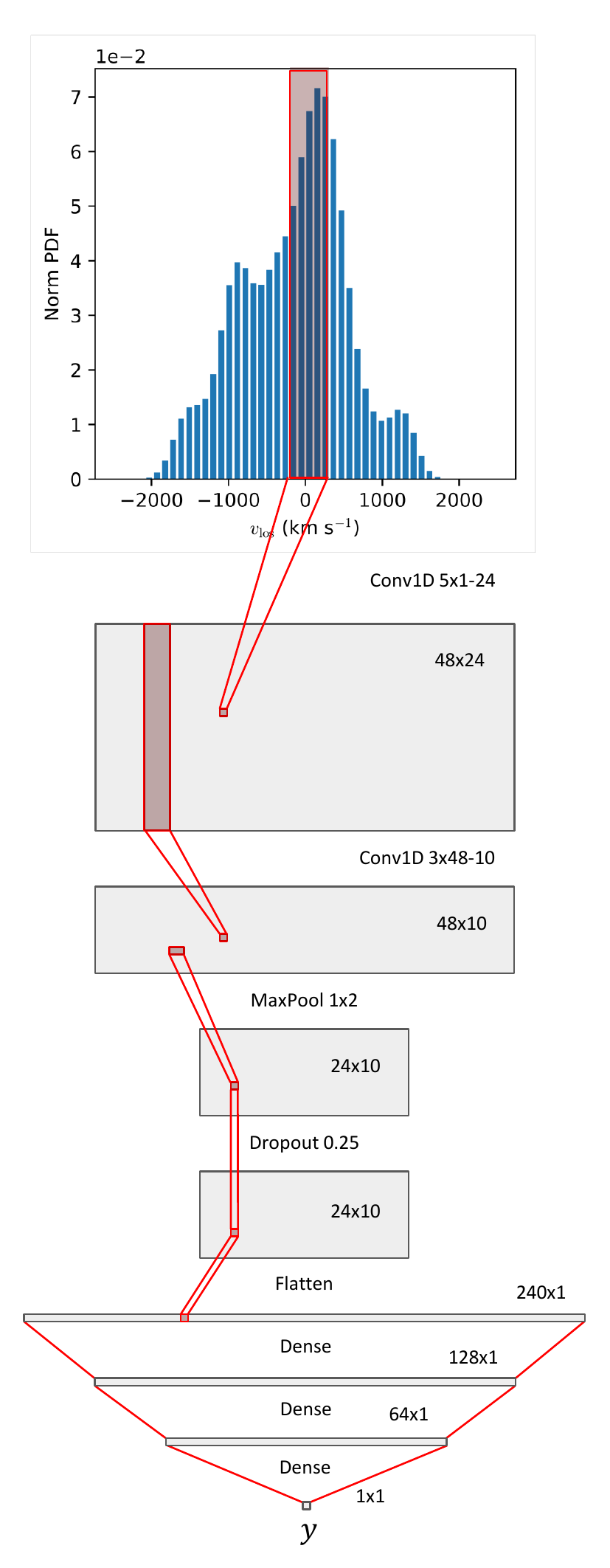}{0.44\textwidth}{(a) CNN$_\text{1D}$ architecture}
              \fig{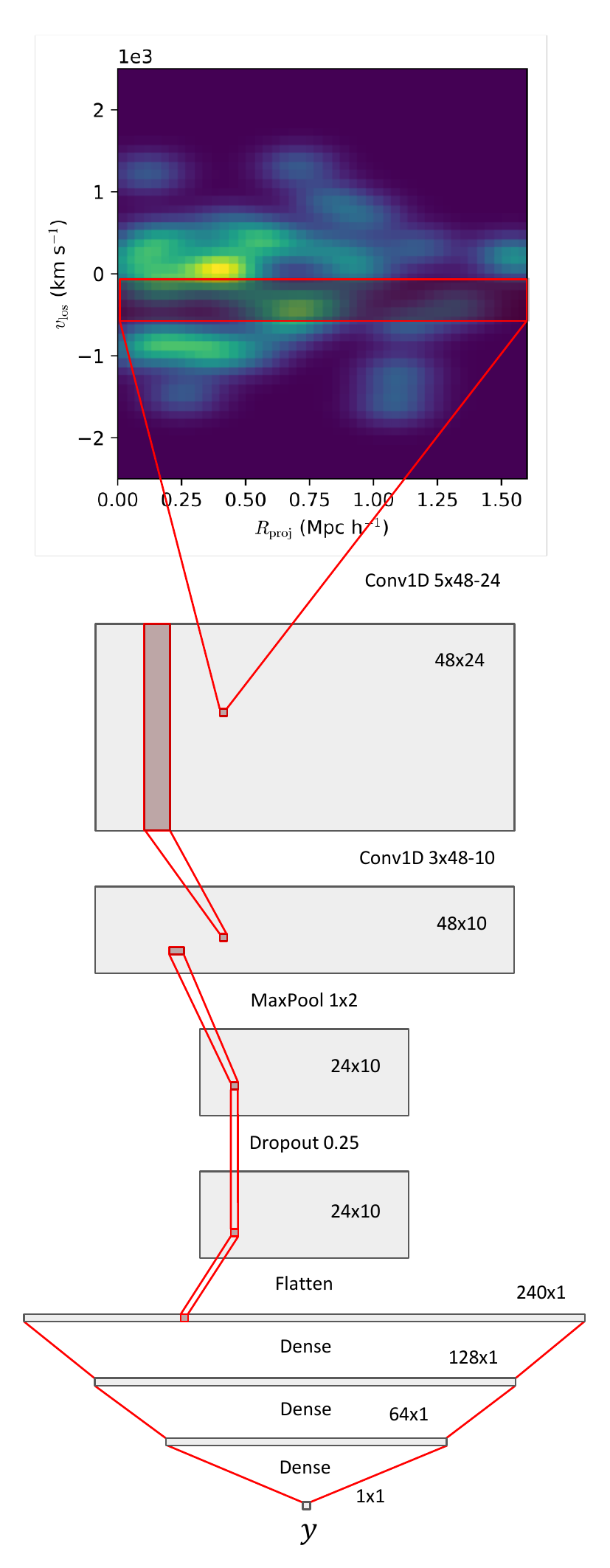}{0.44\textwidth}{(b) CNN$_\text{2D}$ architecture}}
    
    \caption{CNN architecture for each model. The architectures for each case are identical except for the input array and the first convolutional layers. The output of each model $y$ (Equation \ref{eqn:y}) varies linearly with the predicted logarithmic cluster mass and is restricted to the range $y\in[0,1]$. Each layer is subject to a ReLU activation function, and the weight vectors are constrained to a maximum L2 norm of 3.}
    \label{fig:arch}
\end{figure*}

CNNs \citep{LeCun1998} are a subset of DNNs, which mainly benefit from, and are named for, their use of convolutional layers. Unlike dense connections, convolutional connections restrict neurons in one layer to receive information only from neurons within a small neighborhood of the previous layer, called a \textit{receptive field}. This local receptive field method allows neurons to extract simple features from subsets of the input layer, the information from which can be combined to form higher-order features in subsequent layers. The input receptive fields of adjacent neurons within a convolutional layer often overlap, forming a contiguous transformation from input to output, akin to a convolution. The \textit{filter} or \textit{feature extractor}, the set of weights and biases that connect the small region of inputs to the output node, is shared across the entire input layer. This allows the same feature to be detected in different receptive fields across the input while also reducing the complexity of the connection. The output of a filter applied to all regions of an input is called a \textit{feature map}. A full convolutional layer often consists of multiple feature maps, each with different filters. A physical depiction of convolutional layers and their filters can be seen in Figure \ref{fig:arch}.

Convolutional layers within a CNN are often followed by a \textit{pooling layer}. Pooling layers perform a downsampling operation intended to reduce the dimensionality of the convoluted feature maps. The downsampling operation functions in a similar manner to the convolutional filters, in that they execute on local receptive fields across the input. A common downsampling operation is max pooling, in which only the maximum activation from the local receptive field is passed to the next layer.

CNNs, and DNNs in general, use \textit{dropout layers} as a type of stochastic regularization to avoid overfitting. Dropout layers randomly set some fraction of neurons from the previous layers equal to 0 during training. This forces the network to learn feature relationships through multiple neuron paths, reducing training time and preventing overfitting.

Typical simple CNN architectures consist of alternating convolutional and pooling layers followed by several dense layers. Each successive convolutional layer produces coarser, higher-order feature maps of the original input. The final dense layers relate the highest-order features to an output vector. CNNs use the same training procedure as discussed for DNNs.

\subsubsection{Architecture}

The CNN models used in this analysis (Figure \ref{fig:arch}) were designed to incorporate layering patterns common to image-recognition applications while minimizing architectural complexity. Both models use two convolutional layers followed by a max pooling layer, a dropout layer, and three dense layers. The inputs to the models are generated from KDEs as discussed in Section \ref{subsubsec:input}. Each model outputs a single variable $y$, which ranges from $0\leq y\leq 1$ and relates linearly to a mass prediction $\log_{10}[\Hat{M}_\text{pred}\ (h^{-1}\mathrm{M}_\odot)]$.

\begin{equation}
    \log_{10}[\hat{M}_\text{pred}] = \log_{10}[M_\text{min}] + y\log_{10}\Bigg[\frac{M_\text{max}}{M_\text{min}}\Bigg],
    \label{eqn:y}
\end{equation}

where $M_\text{min}$ and $M_\text{max}$ are the minimum and maximum values for $M_{200c}$ in our sample. All masses are expressed in units of $h^{-1}\mathrm{M}_\odot$.   

The convolutional and dense layers in both architectures use a kernel normalization constraint and a rectified linear unit (ReLU) activation function. The kernel constraint normalizes the weighting vector for the input of a given neuron to a constant. The ReLU function is given by the simple form $f(x) = \max(x,0)$. The ReLU activation function has been shown to not saturate as much as conventional sigmoid functions \citep{Nair2010}.

The architectures for CNN$_\text{1D}$ and CNN$_\text{2D}$ are nearly identical, with an exception made for the first convolutional layer. In the first layer, both models use 1-D convolution filters of width 5, which pass over receptive fields along the $v_\text{los}$ axis with a stride of 1. The difference between these architectures is that the CNN$_\text{1D}$ model's filters are of shape $5\times1$ while the CNN$_\text{2D}$ model's filters are of shape $5\times48$. This is done to account for the difference in input shape between the two models. In the first neural layer, both models use 24 filters to create 24 feature maps of length 48. As a result, the outputs of the initial convolutional layers of both CNN$_\text{1D}$ and CNN$_\text{2D}$ are of shape $48\times24$.

\subsection{Training and Evaluation}\label{subsec:training}

\begin{figure}
    \centering
    \includegraphics[width=0.47\textwidth]{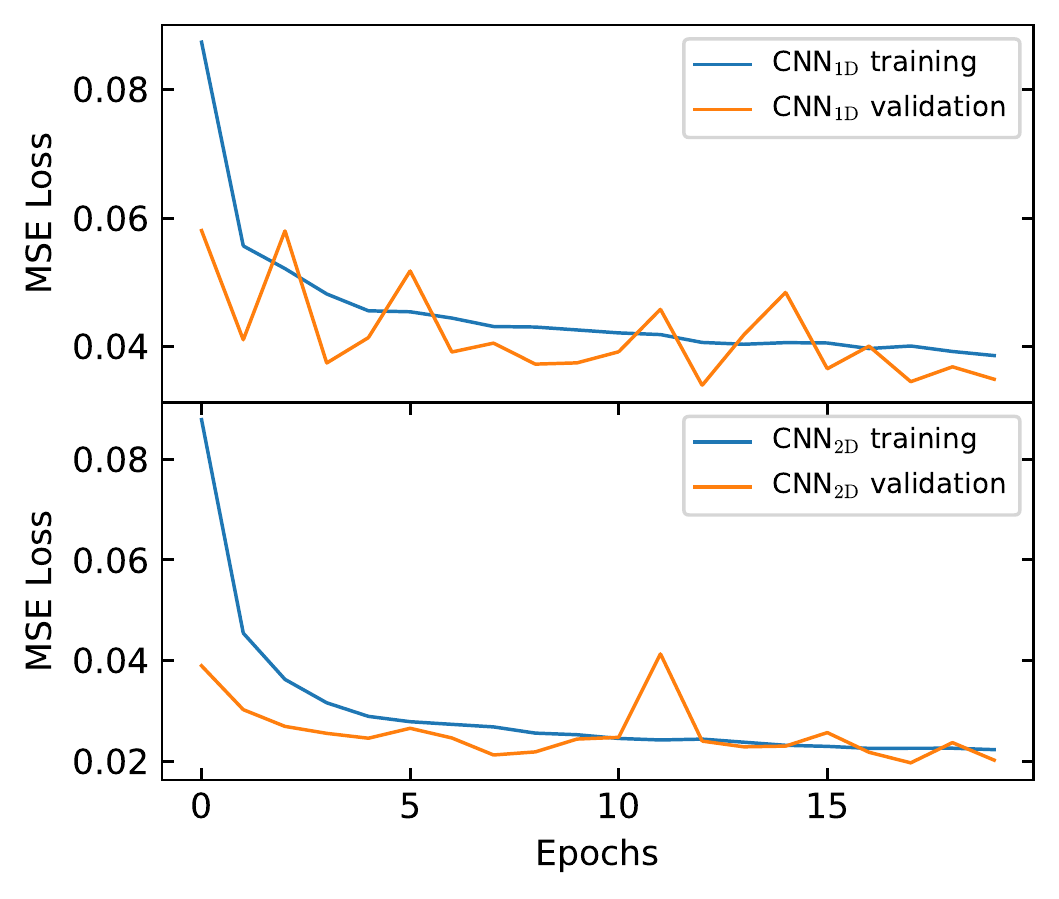}
    \caption{Evolution of mean squared error 
    (MSE) loss for a single fold during CNN$_\text{1D}$ and CNN$_\text{2D}$ training. Training progression exhibits gradual improvement in both the training and validation sets. The validation loss appears less than the training loss due to the introduction of dropout layers.}
	\label{fig:loss}
\end{figure}

We train each model as a regression over the single output variable $y$ (Equation \ref{eqn:y}) using a mean squared error loss function. This is equivalent to minimizing the sum of squared residuals of the output variable $y$, i.e. $\sum\left(y_\text{pred} - y_\text{true}\right)^2$, over the space of our model parameters. For our optimization procedure, we use the Adam protocol \citep{Kingma2014}, a variant of SGD that accounts for momentum and adaptive learning rates in a straightforward, computationally efficient manner. We parameterize the Adam optimizer with a learning rate of $10^{-3}$ and a decay rate of $10^{-6}$. We use a batch size of 100 samples and achieve loss convergence within 20 epochs.

We use a 10-fold cross-validation scheme to evaluate our model. For a given fold, we train on $9/10$ of the cluster candidates in our catalog and test on the remaining, independent $1/10$. This process cycles for 10 folds until predictions have been made for the entire test set. Cluster candidates are grouped along with their rotated LOS duplicates in the training-test split, such that we are never training and testing on the same cluster from different LOSs. This ensures independence of training and testing data for each fold. On average, there are $\sim$10,000 training and $\sim$7,000 test cluster candidates for a given fold.

A validation set is constructed from a disjoint 10\% random sampling of the independent test data. Figure \ref{fig:loss} shows training and validation loss curves for a single fold during the 20 epoch training procedure. The loss curves from both the CNN$_\text{1D}$ and CNN$_\text{2D}$ show gradual improvement throughout training evolution, indicating that neither model is overfitting.

The CNN models and training procedure are implemented using the \textit{Keras}\footnote{\href{https://keras.io/}{https://keras.io/}} library with a \textit{Tensorflow}\footnote{\href{https://www.tensorflow.org/}{https://www.tensorflow.org/}} backend. Each ML analysis was run on two Intel Haswell (E5-2695 v3) CPU nodes with 14 cores and 128 GB of total RAM. The full 10-fold training procedure is executed to convergence in $\sim$10 minutes for both CNN architectures. The KDE generation and sampling process takes $\sim73\mu$s and $\sim410\mu$s per input, for CNN$_\text{1D}$ and CNN$_\text{2D}$ respectively. Once the models are trained and the KDEs are sampled, evaluation time of either CNN neural architecture lasts $\sim$44 $\mu$s per input.

\section{Comparative Methods}\label{sec:compare}
In our comparative analysis, we discuss the performance of CNN$_\text{1D}$ and CNN$_\text{2D}$ relative to other dynamical mass estimation techniques, namely the classical $M$-$\sigma$ and SDM \citep{Ntampaka2015,Ntampaka2016}. Each of these models are evaluated in the context of the mock catalog described in Section \ref{sec:dataset}.

\subsection{$M$-$\sigma$}\label{subsec:msigv}

\begin{figure*}
    \centering
    \gridline{\fig{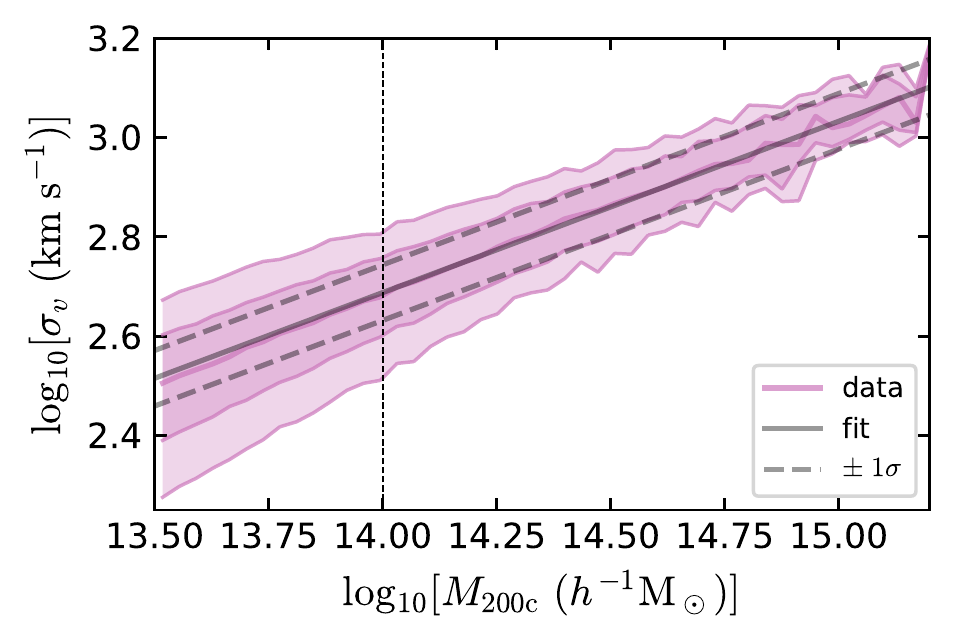}{0.48\textwidth}{(a) Pure $M$-$\sigma$}
              \fig{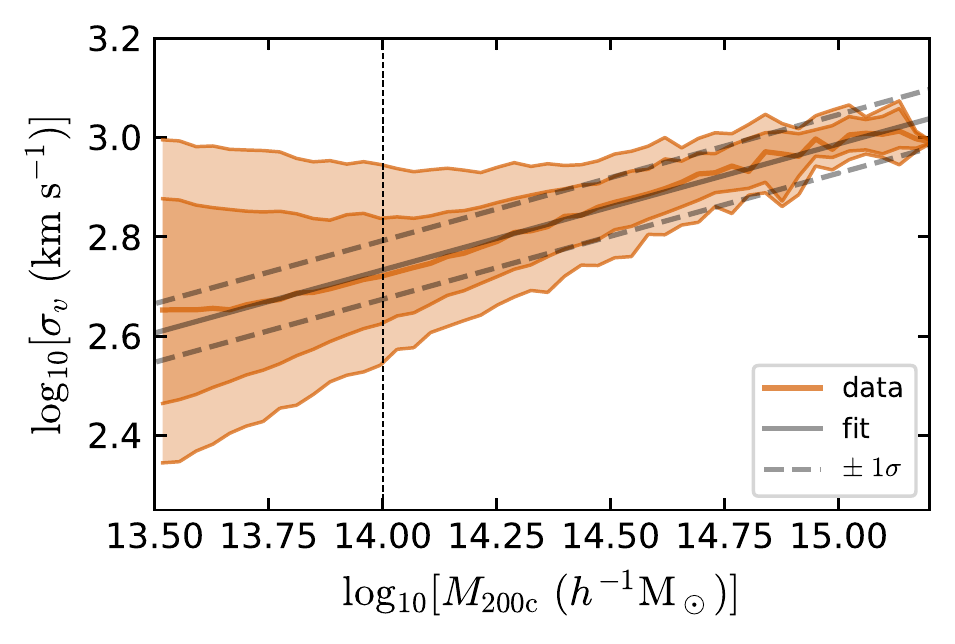}{0.48\textwidth}{(b) Contaminated $M$-$\sigma$}}
    \caption{$M$-$\sigma$ relationship for pure (a) and contaminated (b) mock observation cluster catalogs derived from MDPL2 data. Each distribution is plotted at its median (solid line), 16-84 percentile range (dark region), and 3-97 percentile range (light region). The log-linear regression lines are shown along with their $\pm\ 1\sigma$ lognormal scatter. The dotted black line at $M_{200c} = 10^{14.5}\ h^{-1}\mathrm{M}_\odot$ signifies the lower-bound mass cut used to perform the log-linear regression. Selection effects in the contaminated catalog introduce significant scatter and bias at low masses.}
 	\label{fig:UM_Msigv}
\end{figure*}

The $M$-$\sigma$ scaling relation infers cluster mass from a single summary statistic, the galaxy velocity dispersion $\sigma_v$. If we assume clusters to be stable, spherically symmetric, and purely evolving with gravity, we can derive the classical form of the $M$-$\sigma$ from the kinetic-potential energy equivalence described in the virial theorem. Stated with the appropriate normalization for galaxy clusters, the $M$-$\sigma$ is as follows:
\begin{equation}
    \sigma_v = \sigma_{v,15}\left[\frac{h(z)M_{200c}}{10^{15}\ \mathrm{M}_\odot}\right]^\alpha
    \label{eqn:msig}
\end{equation}
where $M_{200c}$ is our cluster mass definition of a spherical region of density $200\rho_c$, $\sigma_{v,15}$ is a scaling factor parameterizing the velocity dispersion of a galaxy cluster of $M_{200c}=10^{15}\ h^{-1}\mathrm{M}_\odot$, $h(z)$ is the dimensionless Hubble parameter, and $\alpha$ is the power-law scaling parameter. Assuming spherical symmetry, the velocity dispersion $\sigma_v$ can be conveniently taken to be the standard deviation of galaxy velocities projected along a single LOS. The parameter $\alpha$ captures information about the spatial distribution of mass in the spherical cluster and is generally fit with simulation \citep{Evrard2008}.

We perform an $M$-$\sigma$ analysis on both the contaminated catalog described in Section \ref{subsec:contam_mocks} and a comparative, idealized \textit{pure} catalog. Mock observations in the pure catalog are designed to neglect all member selection effects by assuming pure and complete cluster membership. Cluster member samples are constructed from all galaxies which are associated with the cluster's MDPL2 Rockstar FOF group. From this pure member sample, mock observables are calculated in the familiar manner (Section \ref{apx:mock_calc}). The pure cluster catalog is designed to mimic data products of optimal interloper removal strategies, producing a lower limit on $M$-$\sigma$ measurement scatter for modern dynamical mass estimation techniques. Conversely, the cylindrical cuts taken in the contaminated catalog are decidedly simpler than modern methods and thereby produce an upper limit on $M$-$\sigma$ scatter.

We find best-fit parameters $\sigma_{v,15}$ and $\alpha$ for both the pure and contaminated mock catalogs. We use the unbiased standard deviation (Equation \ref{eqn:ustdev}) to estimate velocity dispersions for each cluster sample.

\begin{equation}
    \sigma_v = \sqrt{\frac{1}{N_\text{gal}-1}\sum_{i=1}^{N_\text{gal}}\left(v_\text{los,i} - \bar{v}_\text{los}\right)^2},
    \label{eqn:ustdev}
\end{equation}
where $N_\text{gal}$ is the number of galaxies in a given cluster sample, $v_\text{los,i}$ is the line-of-sight velocity of the $i$-th cluster, and $\bar{v}_\text{los}$ is the average line-of-sight velocity for the cluster. We use an ordinary least-squares linear regression model to fit the power law in log-space: $\log_{10}(\sigma_v) = A\log_{10}(M_{200c}) + B$. As demonstrated in Figure \ref{fig:UM_Msigv}, the contaminated cluster's $M$-$\sigma$ relationship exhibits a departure from log-linear dependence at low masses, due primarily to the saturation of mock observations with unbound galaxies. This is a direct result of the fixed-size cylindrical cuts and was explored in detail in \citet{Ntampaka2016}. When fitting the $M$-$\sigma$, we choose to take a linear regression above a mass cut of $10^{14.5}\ h^{-1}\mathrm{M}_\odot$ and subsequently extrapolate to lower masses. This mass cut is implemented for both the pure and contaminated $M$-$\sigma$ regressions. In addition, both regressions use the flat cluster mass function training set described in Section \ref{subsec:traintest}.

\begin{table}
    \centering
    \begin{tabular}{c|c c|c} \hline
         Catalog & $\sigma_{v,15}\ \left(\mathrm{km}\ \mathrm{s}^{-1}\right)$ & $\alpha$ & Scatter (dex)\\ \hline
         Pure & 1078 & 0.345 & 0.056 \\
         Contaminated & 971 & 0.254 & 0.059 \\ \hline
    \end{tabular}
    \caption{Best-fit parameters for log-linear regression of $M$-$\sigma$ in the pure and contaminated catalogs. Parameters are defined in the formalization of the $M$-$\sigma$ given in Equation (\ref{eqn:msig}). The lognormal scatter is defined as the standard deviation of prediction residuals for clusters above the mass cut, $M_{200c} \geq 10^{14.5}\ h^{-1}\mathrm{M}_\odot$.}
    \label{tab:m_sig}
\end{table}

The $M$-$\sigma$ distribution for pure and contaminated catalogs is shown in Figure \ref{fig:UM_Msigv}. Best-fit parameters are calculated for $\sigma_{v,15}$ and $\alpha$ and are tabulated in Table \ref{tab:m_sig}. We evaluate the lognormal scatter by taking the standard deviation of the residual $\delta$ for clusters above the mass cut,
\begin{equation}
    \delta = \log_{10}\left[\frac{\sigma_{v,\text{pred}}}{\sigma_{v,\text{true}}}\right],
\end{equation}
where $\sigma_{v,\text{true}}$ is the true velocity dispersion for a given cluster and $\sigma_{v,\text{pred}}$ is its predicted velocity dispersion from its true mass and best-fit parameters $\sigma_{v,15}$ and $\alpha$. The parameter values presented in Table \ref{tab:m_sig} are representative of values previously derived from simulation \citep{Evrard2008}, but also exhibit variation due to differences in mock observation strategy.

The $M$-$\sigma$ predictions for both the pure and contaminated catalogs exhibit significant scatter. In the pure case, this scatter can be attributed to physical effects which distort cluster shape or mass distribution. Clusters are highly complex systems in which assumptions of gravitational equilibrium or spherical symmetry are unreliable. In practice, features such as dynamical substructure \citep{Old2018}, halo environment \citep{White2010}, cluster triaxiality \citep{Svensmark2015}, and mergers \citep{Ribeiro2011} act to increase the scatter of $M$-$\sigma$ predictions. In the contaminated case, the prediction scatter is higher than the pure catalog due to the introduction of selection effects \citep{Wojtak2018}. Realistic cluster observations may be incomplete or otherwise contaminated by interloper galaxies. In modern applications of the $M$-$\sigma$, complex membership modeling and interloper removal schemes may be applied to reduce the impact of selection effects \citep[e.g.][]{Wojtak2007, Mamon2013, Farahi2016, Farahi2018, Abdullah2018}, ideally producing predictions equivalent to our pure catalog. Our pure and contaminated predictions therefore define lower and upper bounds, respectively, of the scatter apparent in real $M$-$\sigma$ predictions.

\subsection{Support Distribution Machines}
\textit{Support Distribution Machines} \citep[SDMs;][]{Sutherland2012} are a class of ML algorithms which perform scalar regression over a set of probability distributions. SDMs effectively function as an extension of kernel \textit{support vector machine} \citep[SVM;][]{Scholkopf2002} regression, where nonlinear input is mapped to a space of linear features via some kernel function. Each input to SDM is a variable-length set of i.i.d. samples chosen from an underlying probability distribution. The output is some continuous, scalar value quantifying something about the base probability distribution. SDMs are nonparametric and trained transductively, meaning the complexity of the model is directly proportional to the size of the dataset (train + test). The first application of SDMs to dynamical mass measurements was made in \citet{Ntampaka2015, Ntampaka2016}, where SDMs were used to directly infer cluster mass from lists of galaxy velocities and positions. The SDM approach was effective in reducing $M$-$\sigma$ prediction scatter by a factor of 2. Here, we evaluate SDM performance in the context of our catalog to serve as a baseline with which to compare our ML model.

Replicating our treatment of CNN models, we train SDMs on two types of cluster descriptions, the member $\{v_\text{los}\}$ distribution and the joint member $\{R_\text{proj},v_\text{los}\}$ distribution. We will appropriately refer to these as SDM$_\text{1D}$ and SDM$_\text{2D}$, respectively. Each individual input to the SDM is a list of univariate or bivariate galaxy properties (velocities and/or radial positions). The length of each input list is variable and equal to the cluster richness. In this application of SDMs, we assume this list of galaxies is representative of some underlying probability distribution which varies with cluster mass.

Our implementation of SDM mirrors that of \citet{Ntampaka2016}. The kernel function employed in our SDM model is a Kullback-Leibler divergence, estimated using the $k$-nearest-neighbor method \citep{Wang2009} with $k=3$. We use three-fold cross-validation to find optimal values for SDM parameters $C$ and $\sigma$, the loss function parameter and Gaussian kernel parameter, respectively. We evaluate the SDM models with ten-fold cross-validation, and the training and test sets described in Section \ref{subsec:traintest}.

\begin{table*}[!htb]
    \centering
    \begin{tabular}{c c c c c | r c r c }\hline
         Model & This Paper?  & Data & Color & Catalog &  \multicolumn{1}{c}{$\tilde{\epsilon} \pm \Delta\epsilon$\footnote{Residual median and 16-84 percentile range (dex)}}& 
         $\sigma_\epsilon$\footnote{\label{cite:residual}Residual standard deviation scatter (dex), skewness, and excess kurtosis, respectively}&
         \multicolumn{1}{c}{$\gamma\phantom{}^{\text{\ref{cite:residual}}}$} & 
         $\kappa\phantom{}^{\text{\ref{cite:residual}}}$\\\hline
         CNN$_\text{1D}$& \checkmark & $\{v_\text{los}\}$ & green & Contaminated &  $-0.003^{+0.173}_{-0.163}$ & $0.174$ & $0.419$ & $0.826$\\
         CNN$_\text{2D}$& \checkmark  & $\{R_\text{proj},v_\text{los}\}$ & blue & Contaminated &  $-0.003^{+0.119}_{-0.125}$ & $0.132$ & $0.221$ & $1.600$\\
         M-$\sigma_\text{pure}$ & & $\{v_\text{los}\}$ & violet & Pure &  $0.006^{+0.179}_{-0.195}$ & $0.193$ & $-0.262$ & $0.417$\\
         M-$\sigma_\text{contam}$ & & $\{v_\text{los}\}$ & orange & Contaminated &  $-0.016^{+0.300}_{-0.290}$ & $0.316$ & $0.225$ & $0.647$\\
         SDM$_\text{1D}$\footnote{\label{cite:ntampaka}\citet{Ntampaka2016}} & & $\{v_\text{los}\}$ & yellow & Contaminated &  $-0.039^{+0.229}_{-0.187}$ & $0.226$ & $0.646$ & $1.183$\\
         SDM$_\text{2D}\phantom{}^{\text{\ref{cite:ntampaka}}}$& & $\{R_\text{proj},v_\text{los}\}$ &  pink & Contaminated & $-0.018^{+0.149}_{-0.148}$ & $0.159$ & $0.309$ & $1.459$\\
        \hline
    \end{tabular}
    \caption{Summary of investigated models. In addition to the CNN models presented in this paper, we include other comparative dynamical mass estimates, including the traditional $M$-$\sigma$ and a modern ML approach \citep[SDM;][]{Ntampaka2016}. We analyze the $M$-$\sigma$ method under both a pure and contaminated catalog in order to provide lower and upper bounds on the scatter of general interloper removal strategies. We include several cumulative statistics describing the error (Section \ref{subsec:perf}) and lognormality (Section \ref{subsec:lognorm}) of each model's mass predictions.}
    \label{tab:models}
\end{table*}

Analysis of each SDM model was run on two Intel Haswell (E5-2695 v3) CPU nodes with 14 cores each and 128 GB of total RAM. Using the mock catalog described in Section \ref{sec:dataset}, the full 10-fold transductive training and evaluation procedure executed in $\sim$6 hr for each SDM model.

\section{Results} \label{sec:results}

The results presented in this section analyze the performance of our CNN models when evaluated on a catalog of mock cluster observations (Section \ref{sec:dataset}). Model performance is quantified in terms of predictive scatter, bias, log-normality, robustness, and application time. We describe these metrics in the context of observational studies and discuss their implications in precision cosmology. Using these metrics, we perform comparative analyses with respect to the dynamical mass estimators described in Section \ref{sec:compare}. The complete list of investigated models presented in this section is summarized in Table \ref{tab:models}. We find that the CNN models produce more accurate and robust mass estimates than all other investigated methods, with considerably shorter implementation times than SDM.

\subsection{Predictive Performance} \label{subsec:perf}

\begin{figure*}
    \centering
    \gridline{\fig{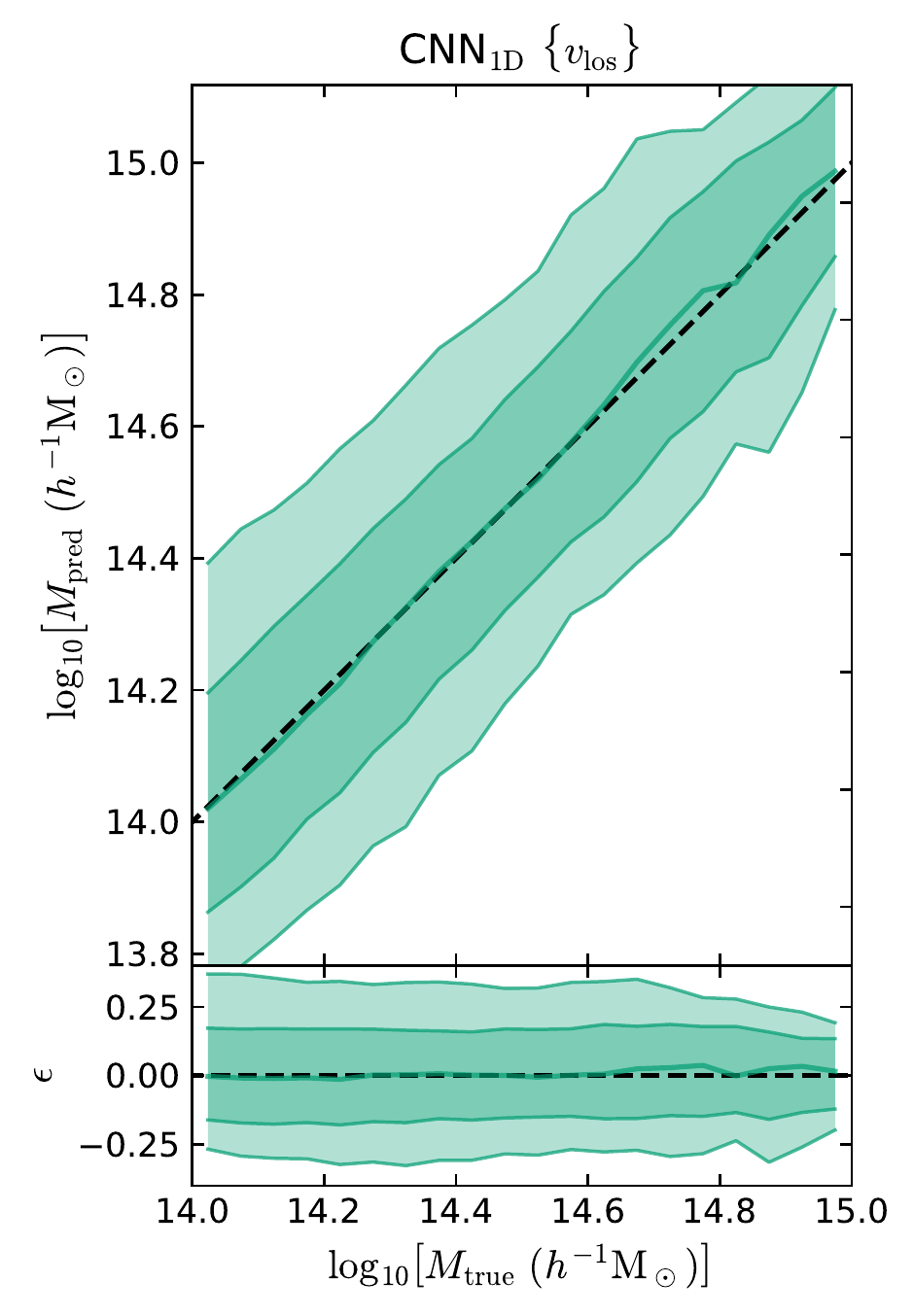}{0.49\textwidth}{(a) CNN$_\text{1D}$ Predictive Performance}
              \fig{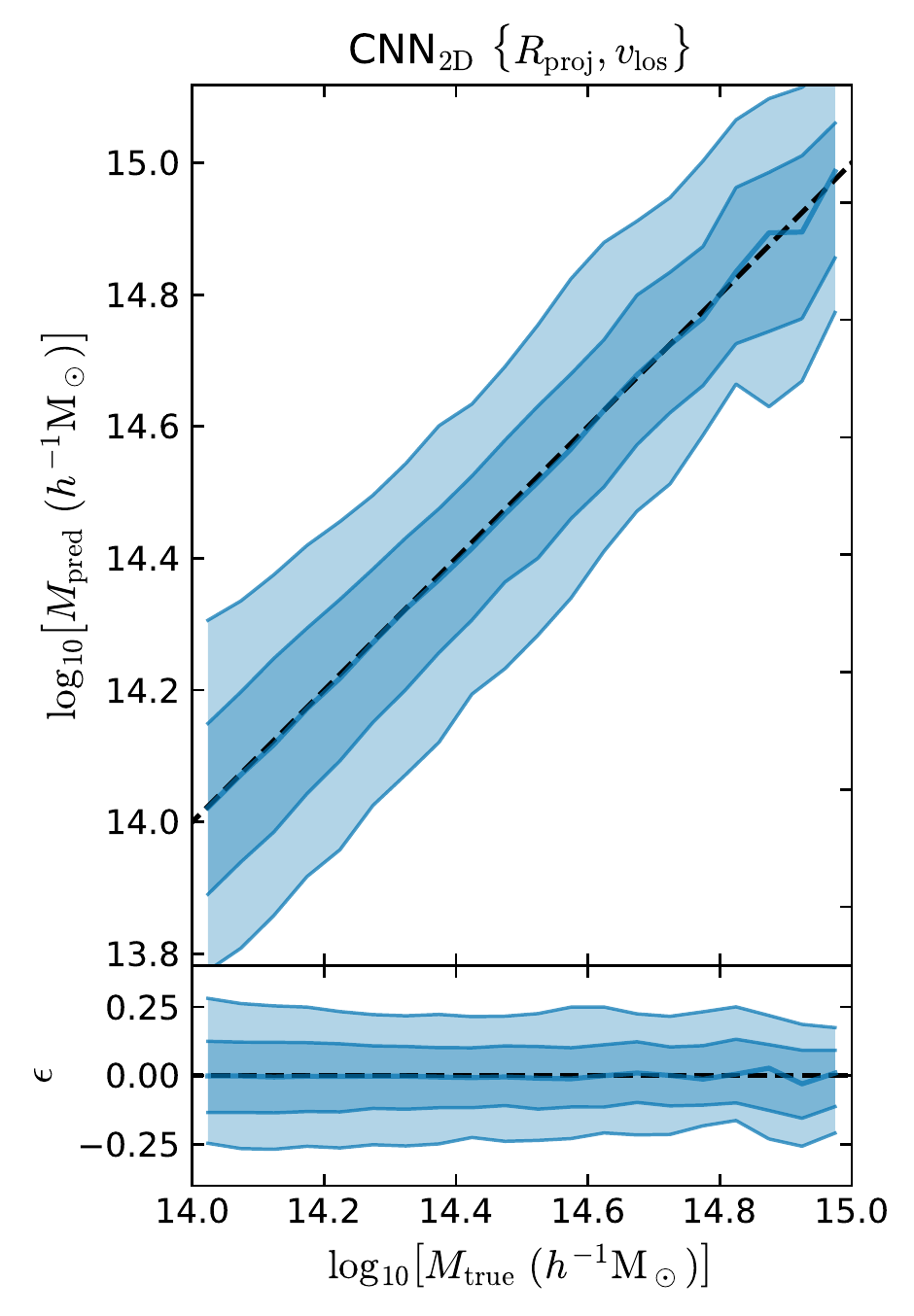}{0.49\textwidth}{(b) CNN$_\text{2D}$ Predictive Performance}}
	\caption{Predicted-vs.-true mass distributions for our CNN models when predicting a realistic sample of mock cluster observations. Panel (a) shows the binned distribution of predicted masses and residuals (Equation \ref{eqn:err}) using CNN$_\text{1D}$. Each distribution is plotted at its median (solid line), 16th-84th percentile range (dark region), and 3rd-97th percentile range (light region). Panel (b) shows the same prediction and residual distributions for CNN$_\text{2D}$. The mass definition applied in this analysis is $M_\text{true}=M_\text{200c}$.}
	\label{fig:perf}
\end{figure*}

\begin{figure*}
    \centering
    \includegraphics[width=\textwidth]{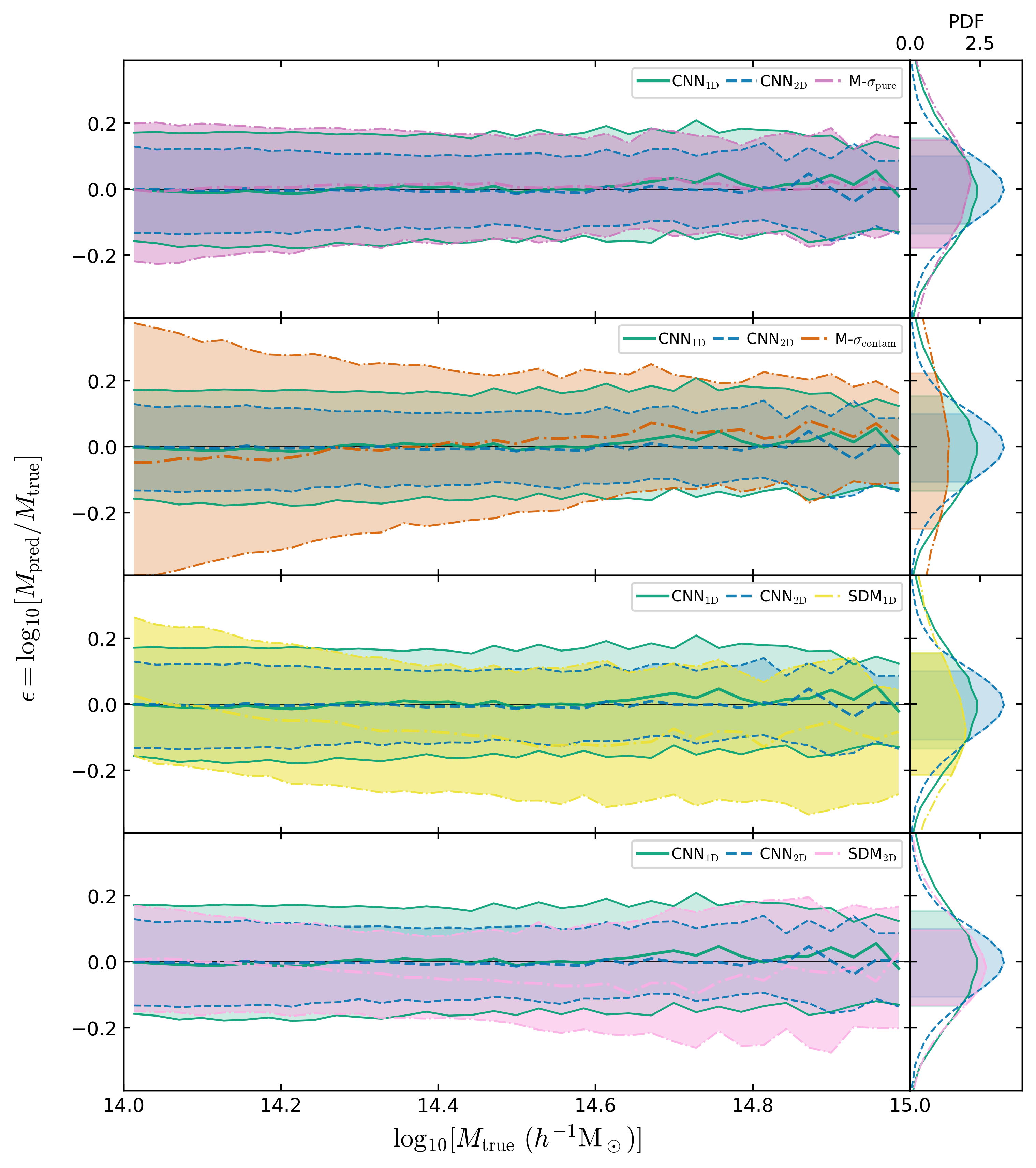}
    \caption{Prediction residuals $\epsilon$ (Equation \ref{eqn:err}) for CNN$_\text{1D}$ and CNN$_\text{2D}$ relative to comparative models (Table \ref{tab:models}), including the traditional $M$-$\sigma$ and a modern ML approach \citep[SDM;][]{Ntampaka2016}. For clarity, comparisons with various models are shown on separate rows, in the order of Table \ref{tab:models}. \textit{Left column}: residual distributions are binned along true mass and shown at their median and 16-84 percentile range. \textit{Right column}: residual distributions marginalized over true mass and plotted as PDFs. The highlighted region corresponds to the marginalized 16th-84th percentile range.}
	\label{fig:err}
\end{figure*}

Figure \ref{fig:perf} shows the multi-fold predicted-versus-true mass distribution of the CNN$_\text{1D}$ and CNN$_\text{2D}$ models when performing inference on the test data set (Section \ref{subsec:traintest}). For each model, we describe the distribution of mass predictions via the logarithmic residual $\epsilon$, defined as
\begin{equation}
    \epsilon = \log_{10}\left[\frac{M_\text{pred}}{ M_\text{true}} \right]
    \label{eqn:err}
\end{equation}
for a cluster of mass $M_\text{true}$ whose predicted mass is $M_\text{pred}$. This metric is commonly employed in other observational studies \citep[e.g.][]{Armitage2018, Armitage2019, Calderon2019} and conveniently scales linearly with our model output $y$ (Equation \ref{eqn:y}). The mass definition used in this analysis is $M_\text{true}=M_\text{200c}$. We further characterize model predictions by calculating cumulative statistics of the $\epsilon$ distribution, namely the median ($\tilde \epsilon$), 16th-84th percentile range ($\Delta\epsilon$), and the standard deviation scatter ($\sigma_\epsilon$). The values of these statistics for CNN$_\text{1D}$ and CNN$_\text{2D}$ are tabulated in Table \ref{tab:models}. Note that these cumulative statistics are constructed from the test catalog and marginalized over true mass and are thereby weighted by the shape of the test catalog cluster mass function (Figure \ref{fig:hmf}). 

As seen in Figure \ref{fig:perf}, the CNN model predictions exhibit low scatter and bias across the test mass range. The residual scatter $\sigma_\epsilon$ for CNN$_\text{2D}$ predictions, $0.132$ dex ($\simeq 30\%$), is considerably lower than that for CNN$_\text{1D}$ predictions, $0.174$ dex ($\simeq 40\%$), indicating that the supplementary information about underlying galaxy distributions provided by $R_\text{proj}$ reduces scatter by $24\%$ under the CNN framework. Each model's $\epsilon$ distribution shows a marginal trend toward higher scatter at low true mass, which we attribute to a reduction of true members and a saturation of interlopers in the fixed cylindrical membership cut.

Figure \ref{fig:err} plots the median and 16-84 percentile range of prediction residuals as a function of true mass for each investigated model listed in Table \ref{tab:models}. Each model is evaluated on the same contaminated mock catalog (Section \ref{sec:dataset}) except for M-$\sigma_\text{pure}$ which is evaluated on a catalog with perfect membership selection (Section \ref{subsec:msigv}). The SDM and $M$-$\sigma$ models serve as baselines for modern ML and interloper removal schemes, respectively. Cumulative statistics for these comparative methods are listed in Table \ref{tab:models}. The prediction scatter measured for $M$-$\sigma$ and SDM methods is consistent with literature \citep{Evrard2008, Ntampaka2016}.

CNN models produce the equivalent or better predictive performance than either pure or contaminated $M$-$\sigma$ measurements. The simple M-$\sigma_\text{contam}$ model exhibits high bias and scatter, with exceptionally high deviation at low masses, resulting from interloper saturation. The $\epsilon$ distribution of CNN$_\text{1D}$ is virtually equivalent to that of M-$\sigma_\text{pure}$, suggesting that CNN$_\text{1D}$ is capable of achieving the same scatter as optimal interloper removal algorithms. Whereas M-$\sigma_\text{pure}$ improves upon M-$\sigma_\text{contam}$ by eliminating selection systematics, the prediction improvements made by CNN$_\text{1D}$ likely stem from a mitigation of both selection and physical effects.  CNN$_\text{2D}$'s low scatter and bias relative to the pure and contaminated $M$-$\sigma$ can be attributed to its use of $R_\text{proj}$ information. These results imply that the CNN models presented here may be preferable over modern $M$-$\sigma$-based interloper removal methods.

According to Table \ref{tab:models}, the SDM$_\text{1D}$ and SDM$_\text{2D}$ models are effective in reducing prediction scatter to below that of M-$\sigma_\text{pure}$, but produce strong prediction biases. Both SDM models observe significant deviations in median prediction $\tilde{\epsilon}$ at various regions in the testing mass range. This is visible in Figure \ref{fig:err}, where SDM$_\text{1D}$ and SDM$_\text{2D}$ underpredict medium- to high-mass clusters. This behavior may complicate applications in precision cosmology. The SDM biases measured here are consistent with results shown in \citet{Ntampaka2016}. Aside from these biases, both SDM$_\text{1D}$ and SDM$_\text{2D}$ produce lower prediction scatter $\sigma_\epsilon$ than CNN$_\text{1D}$. This outcome is intuitive, considering that the KDE step in the CNN approach ``smooths out" distribution information which is potentially informative of cluster mass. However, CNN$_\text{2D}$ is capable of overcoming this hindrance to produce a prediction scatter that is lower than both SDM models. The improved complexity of CNN$_\text{2D}$ is therefore capable of capturing mass-dependent features of cluster dynamics at least as well as applications of SDM.


CNN$_\text{1D}$ and CNN$_\text{2D}$ reduce the prediction scatter $\sigma_\epsilon$ of the contaminated $M$-$\sigma$ measurements by $45\%$ and $58\%$. When compared to the idealized $M$-$\sigma$, these models show $10\%$ and $32\%$ improvement respectively. CNN$_\text{2D}$ shows lower scatter than the best SDM model, producing $17\%$ lower scatter than SDM$_\text{2D}$. The prediction improvements of CNN are comparable to those noted in other ML approaches \citep[e.g.][]{Armitage2018, Calderon2019}. This analysis suggests that CNN$_\text{1D}$ and CNN$_\text{2D}$ are capable of capturing mass-dependent input features and are effective models of cluster dynamics distributions. Under the assumptions made by the simulated catalog listed in Section \ref{subsec:contam_mocks}, CNN$_\text{2D}$ is the most accurate predictor of dynamical cluster masses among the above investigated models.

\subsection{Lognormality} \label{subsec:lognorm}

Mass estimators with non-Gaussian prediction likelihoods can introduce bias in cosmological analyses based on cluster counts \citep{Erickson2011, Weinberg2013}.
We seek to characterize the non-Gaussianity of predictions made by CNN and other comparative methods in order to estimate their impact on halo abundance calculations. We follow a formalism introduced by \citet{Shaw2010} whereby we model the observable-mass relation for a fixed redshift by an Edgeworth expansion,\begin{equation}
    P\left(M_\text{pred}|M_\text{true}\right) \approx G(x) - \frac{\gamma}{6}\frac{d^3 G}{dx^3} + \frac{\kappa}{24}\frac{d^4G}{dx^4},
    \label{eqn:mass-obs}
\end{equation}
where $x = \left(\epsilon-\langle \epsilon \rangle\right)/\sigma_\epsilon$ is the normalized logarithmic residual, $G$ is the standard normal distribution, and $\gamma$ and $\kappa$ are the skewness and excess kurtosis of the $x$ distribution, respectively. For a power-law mass function $[dn/d\ln M]\propto M^{-\alpha}$, cluster abundance measurements can be expressed as 
\begin{equation}
\begin{split}
    \frac{dn}{d\ln M_\text{pred}} \approx &\left(\frac{dn}{d\ln M_\text{pred}}\right)_0\\&
    \times\left[1 + \frac{\alpha^3\sigma^3}{6}\gamma + \frac{\alpha^4\sigma^4}{24}\kappa\right],
    \label{eqn:abundance}
\end{split}
\end{equation}
where $M_\text{pred}$ is defined in terms of $h^{-1}\mathrm{M}_\odot$, $\sigma$ is the logarithmic prediction scatter (in percent), and $(dn/d\ln M_\text{pred})_0$ is the abundance for a purely log-normal $x$ distribution \citep{Weinberg2013}. From Equation \ref{eqn:abundance}, we can estimate the systematic uncertainty in cluster abundance measurements from the mass estimator cumulants $\sigma$, $\gamma$, and $\kappa$.

\begin{figure}
    \centering
    \includegraphics[width=\linewidth]{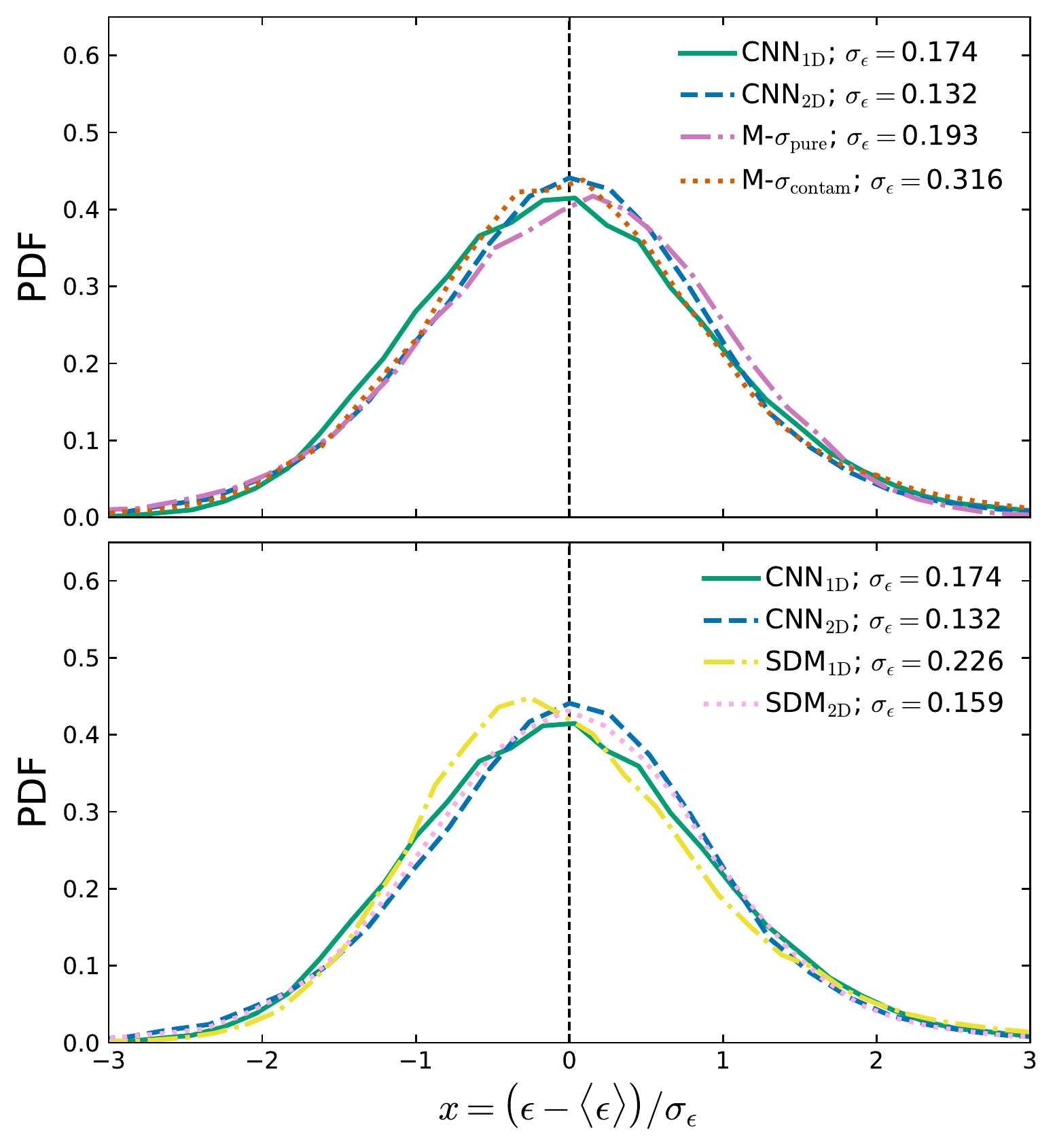}
    \caption{Distribution of normalized prediction residuals marginalized over true mass for each investigated model (Table \ref{tab:models}). Each subfigure plots the PDF of residuals normalized by their mean $\langle\epsilon\rangle$ and scatter $\sigma_\epsilon$. For context, residual scatter $\sigma_\epsilon$ (in dex) is listed in the legend for each model. For clarity, model comparisons with $M$-$\sigma$ (upper) and SDM (lower) are shown on separate plots.}
	\label{fig:lognorm}
\end{figure}

Table \ref{tab:models} lists the lognormality descriptors for each model's mass predictions. Figure \ref{fig:lognorm} draws the PDF of the normalized residual distribution for each investigated model. From these statistics, we see that the PDF of each model's prediction residuals is roughly Gaussian. For a typical power-law mass distribution of slope $\alpha=2$, the impact of non-Gaussian uncertainty on abundance measurements is $\leq5\%$ for all models except M-$\sigma_\text{contam}$ and SDM$_\text{1D}$. M-$\sigma_\text{contam}$'s high systematic uncertainty ($23\%$) is primarily driven by its large scatter $\sigma_\epsilon=0.316$ dex. SDM$_\text{1D}$'s uncertainty ($17.8\%)$ is a result of its biased high mass cluster predictions and resulting residual skewness ($\gamma=0.646$). CNN$_\text{1D}$ predictions produce a low systematic uncertainty of $5.0\%$. CNN$_\text{2D}$ produces the lowest non-Gaussian systematic uncertainty of all investigated models at $1.7\%$, slightly below that of the idealized M-$\sigma_\text{pure}$ at $2.0\%$ and SDM$_\text{2D}$ at $3.8\%$.\\

\subsection{Mass and Richness Dependence}

We adopt the formalism introduced in \citet{Wojtak2018} to characterize the dependence of our models' bias and scatter on cluster mass and richness. Following this formalism, we assume that the distribution of our residuals $\epsilon$ (Equation \ref{eqn:err}) is Gaussian with mass-dependent mean $\mu$ and richness-dependent scatter $\sigma$. We describe our residual distribution according to the following likelihood:
\begin{equation}
    L\propto\prod_i \left[\left(1-w_c\right)G\left(\epsilon_i;\mu,\sigma\right) + w_cG\left(\epsilon_i;\mu,\sigma_c)\right)\right],\label{eqn:likelih}
\end{equation}
where $G\left(\epsilon; \mu,\sigma\right)$ is a Gaussian function of $\epsilon$ with mean $\mu$ and variance $\sigma^2$ and where the product is over the full contaminated catalog test set. The second term in the likelihood accounts for a flat distribution of outliers and is parameterized by the nuisance parameter $w_c$ and scatter $\sigma_c$, the latter of which is fixed to a large value, $\sigma_c=10^{10}\ \mathrm{dex}$. The mean $\mu$ of our residual distribution is modeled as a linear function of logarithmic cluster mass:
\begin{equation}
    \mu = \mu_0 + \left(\alpha_0 - 1\right)\log_{10}\left(M_\mathrm{true}/M_0\right),
\end{equation}
where $\mu_0$ and $\alpha_0$ are free parameters, and the pivot mass $M_0$ is fixed to the median of our cluster sample, $M_0 = 10^{14.17}\ h^{-1}\mathrm{M}_\odot$. The residual scatter, $\sigma$, is related to cluster richness through the following parameterization:
\begin{equation}
    \sigma^2 = \sigma_0^2 + \left(\frac{100}{N_\mathrm{true}}\right)\sigma_1^2,
\end{equation}
where $\sigma_0$ and $\sigma_1$ are free parameters describing the intrinsic and richness-dependent scatter, respectively, and $N_\mathrm{true}$ denotes the true cluster richness as reported by the UniverseMachine catalog, ignoring sample contamination and incompleteness.

We use a Metropolis-Hastings algorithm to sample the likelihood (Equation \ref{eqn:likelih}) and report the best-fit values in Table \ref{tab:mcmc}, marginalizing over the nuisance parameter $w_c$. For both models, our results indicate mass biases consistent with $\mu_0=0$ and a well-constrained log-linear $M_\mathrm{pred}$-$M_\mathrm{true}$ relation ($\alpha_0=1$). As expected, residual scatter for each model scales with cluster richness, with higher cluster richness (more information) leading to a reduced residual scatter. At our median richness of $\tilde{N}_\mathrm{true}=40$, about $55\%$ and $47\%$ of residual scatter can be explained by the intrinsic scatter for CNN$_\text{1D}$ and CNN$_\text{2D}$, respectively.

The mass and richness dependencies of CNN models are comparable to those of the 25 commonly used cluster mass estimation techniques analyzed as part of the Galaxy Cluster Mass Reconstruction Project \citep[GCMRP;][]{Wojtak2018}. Table \ref{tab:mcmc} indicates that the log-linear $M_\mathrm{pred}$-$M_\mathrm{true}$ relations recovered by CNN models show the least intrinsic and mass-dependent biases of all values reported by the GCMRP. In addition, our results suggest that the richness-corrected scatters of CNN models are among the lowest of all GCMRP models. However, we caution the reader against using the values listed in Table \ref{tab:mcmc} as a direct, rigorous comparison with those published in \citeauthor{Wojtak2018}, on account of differences between our dataset and that of the GCMRP. Most notably, the catalogs of mock clusters used in the GCMRP analyses \citep{Old2015} were populated with galaxies using halo occupation distribution (HOD) or semi-analytic models (SAMs), each of which employs procedures considerably different from UniverseMachine. A future joint analysis of the CNN and GCMRP models conducted on the same dataset would provide more reliable comparisons.

\begin{table}
    \centering
    \begin{tabular}{c|c c c c} \hline
         Model & $\mu_0$ & $\sigma_0$ & $\sigma_1$ & $\alpha_0$ \\ \hline
         CNN$_\text{1D}$ & $0.01_{-0.02}^{+0.03}$ & $0.15_{-0.00}^{+0.00}$ & $0.05_{-0.00}^{+0.00}$ & $1.00_{-0.00}^{+0.00}$\\
         CNN$_\text{2D}$ & $0.03_{-0.02}^{+0.03}$ & $0.11_{-0.00}^{+0.00}$ & $0.05_{-0.00}^{+0.00}$ & $1.00_{-0.00}^{+0.00}$\\ 
         \hline
    \end{tabular}
    \caption{Best-fit model parameters characterizing the dependence of prediction residuals on cluster mass and richness \citep{Wojtak2018}. Each entry shows the median and 16-84 percentile range for a Metropolis-Hastings sampling over the parameter space of the likelihood given in Equation \ref{eqn:likelih}.}
    \label{tab:mcmc}
\end{table} 
\begin{figure*}
    \centering
    \includegraphics[width=\linewidth]{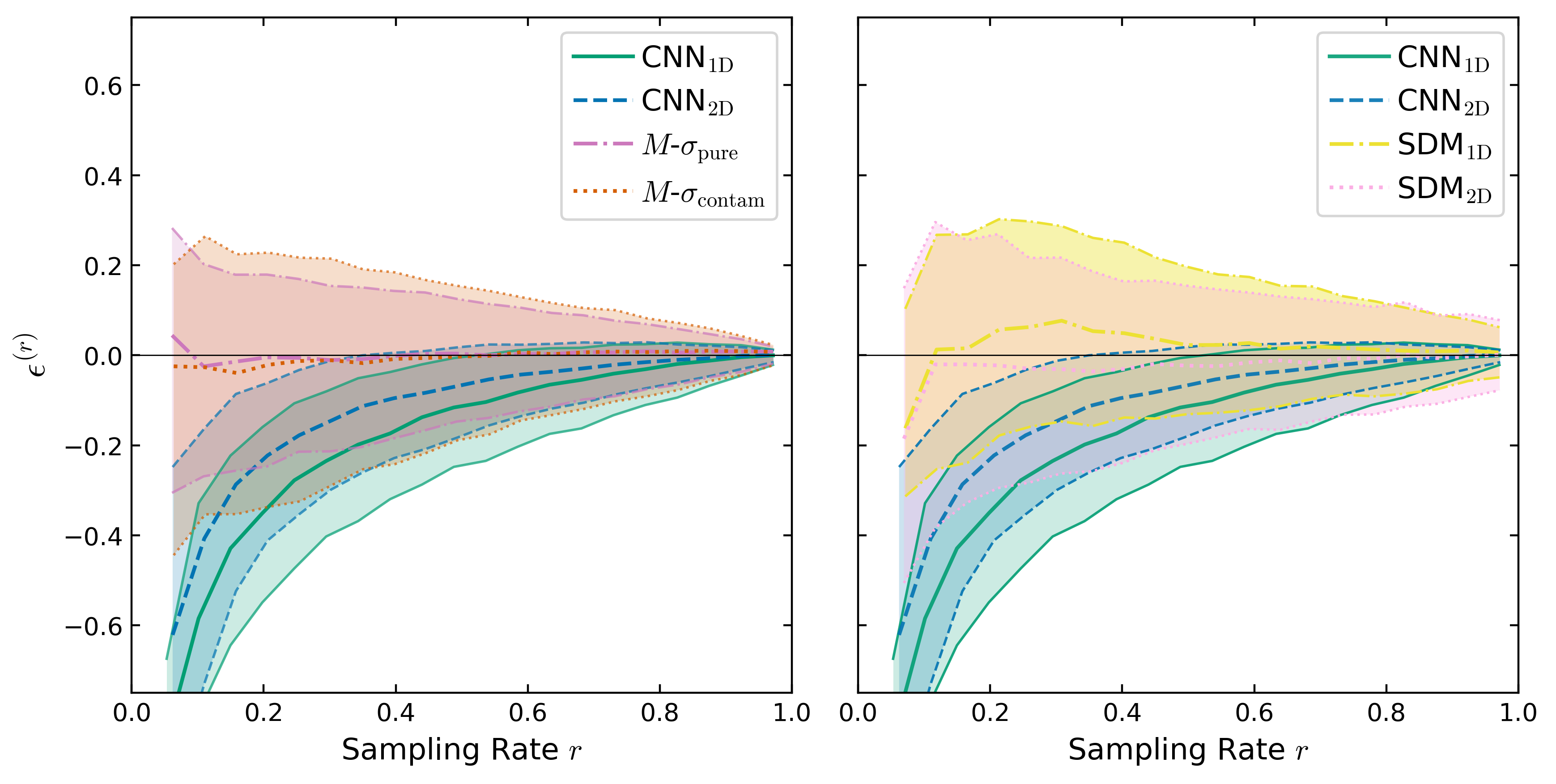}
    \caption{Subsampled mass deviation $\epsilon^{(r)}$ (Equation \ref{eqn:err_subsample}) at a range of sampling rates $0\leq r\leq1$ for CNN$_\text{1D}$, CNN$_\text{2D}$, and comparative models (Table \ref{tab:models}). Subsampled mass deviation is a measure of how model predictions ``drift" when galaxies are randomly removed from the input. The CNN models as plotted here show low prediction drift under variations in galaxy sampling rate relative to other models. These deviation trends are independent of original cluster mass and richness. Distributions are binned and shown at their median and 16th-84th percentile range. For clarity, model comparisons with $M$-$\sigma$ (left) and SDM (right) are shown on separate plots.}
	\label{fig:rich}
\end{figure*}

\subsection{Sampling Variation}

We seek to quantify the robustness of our model predictions under variations in galaxy sampling rate. In practice, this is a measure of the reliability of our mass estimates when some fraction of galaxies are indistinguishable or otherwise not spectroscopically observed, as is common in astronomical observations. We construct \textit{subsampled mass deviation} $\epsilon^{(r)}$ as a measurement of prediction stability. For each model, we define $M^{(r)}_\text{pred}$ as the mass prediction for a given cluster when its set of member galaxies is randomly subsampled at a rate of $r$ without replacement. We choose to subsample randomly so as not to introduce new selection effects. The number of possible subsampled galaxy combinations can be intractably high, so we use the average subsampled mass prediction $\bar{M}_\text{pred}^{(r)}$, calculated from a fixed number of subsampled combinations. For each cluster, we average mass predictions from ten different galaxy subsamplings to assign a single measurement of $\bar{M}_\text{pred}^{(r)}$. Following from this definition, subsampled mass deviation $\epsilon^{(r)}$ (Equation \ref{eqn:err_subsample}) is the logarithmic difference between the average subsampled mass prediction $\bar{M}_\text{pred}^{(r)}$ and the fully sampled prediction $M^{(1.0)}_\text{pred}$. \begin{equation}
    \epsilon^{(r)} = \log_{10}\left[\frac{\bar{M}_\text{pred}^{(r)}}{M_\text{pred}^{(1.0)}}\right].
    \label{eqn:err_subsample}
\end{equation}

The subsampled mass deviation measures how much a model's predictions `drift' on average under fluctuations in sampling rate. Mass measurements that have a high reliance on cluster richness will show a strong correlation between $r$ and $\epsilon^{(r)}$. While accurate, these models may fail when the sampling rate is not well constrained. We construct a cumulative statistic $\tilde{\epsilon}^{(6\text{-}8)} \pm \Delta \epsilon^{(6\text{-}8)}$ which describes the median and 16th-84th percentile scatter of all $\epsilon^{(r)}$ measurements within $0.6\leq r \leq 0.8$. This measurement aims to characterize the bias and scatter involved with using each of the investigated models when the sampling rate is allowed to vary uniformly between 60-80\%. In doing so, we capture the effects of both intrinsic scatter and richness dependence within our models' predictions. Ideal model performance involves producing low values of $|\tilde{\epsilon}^{(6\text{-}8)}|$ and $\Delta \epsilon^{(6\text{-}8)}$. Regardless of model performance, we expect sampling mass fraction to deviate strongly from $\epsilon^{(r)}=0$ as $r\rightarrow0^+$ due to loss of input information.

\begin{table}
    \centering
    \begin{tabular}{c c  | r c  }\hline
         Model &  Color &  \multicolumn{1}{c}{$\tilde{\epsilon}^{(6\text{-}8)} \pm \Delta \epsilon^{(6\text{-}8)} $\footnote{Subsampled mass deviation median and 16-84 percentile range marginalized over sampling rates within $0.6\leq r\leq0.8$ in dex}} & 
         $\Delta \epsilon^{(6\text{-}8)}$ \footnote{16-84 percentile width in dex}\\\hline
         CNN$_\text{1D}$ &  green & $-0.049^{+0.069}_{-0.100}$ & $0.168$\\
         CNN$_\text{2D}$ & blue & $-0.026^{+0.053}_{-0.073}$ & $0.126$ \\
         M-$\sigma_\text{pure}$  & violet & $0.005^{+0.077}_{-0.102}$ & $0.179$\\
         M-$\sigma_\text{contam}$ & orange & $0.006^{+0.095}_{-0.121}$ & $0.216$\\
         SDM$_\text{1D}$\footnote{\label{cite:ntampaka2}\citet{Ntampaka2016}} &  yellow  & $0.017^{+0.126}_{-0.118}$ & $0.244$\\
         SDM$_\text{2D}\phantom{}^{\text{\ref{cite:ntampaka2}}}$ &  pink &  $-0.013^{+0.133}_{-0.135}$ & $0.268$\\
        \hline
    \end{tabular}
    \caption{Cumulative statistics of robustness measurements for all investigated models listed in Table \ref{tab:models}.}
    \label{tab:robust}
\end{table}

Measurements of $\epsilon^{(r)}$ for the CNN and $M$-$\sigma$ models are constructed via inductive learning, i.e., by optimizing model parameters on fully sampled training data and subsequently inferring masses for subsampled test data. Due to the transductive nature of SDM, both train and test data have an impact on SDM model fitting and must be used jointly in the learning procedure. We fit numerous iterations of SDM models, each trained on the same fully sampled training data and evaluated on a unique set of sampled test data. For a single iteration, each cluster in the test data set is subsampled by the same fraction $r$. This is done to mimic realistic observation conditions; each iteration corresponds to observation conditions where the galaxy observation rate is fixed at $r$. Consolidating the mass predictions made by each SDM iteration and comparing them to the fully-sampled $r=1$ predictions produce estimations of $\epsilon^{(r)}$ for the range of possible sampling rates.

Figure \ref{fig:rich} shows the subsampled mass deviation distribution for the investigated ML models and the traditional $M$-$\sigma$ as a function of sampling rate $r$. These sampling variation trends are independent of true sample richness or mass. Cumulative statistics for these distributions are calculated in Table $\ref{tab:robust}$. As expected, each model tends to deviate strongly from $\epsilon^{(r)}=0$ at low $r$ due to loss of input information. At sampling rates $r<0.2$, we see sharp changes in $\epsilon^{(r)}$, suggesting that $r=0.2$ marks a considerable loss of cluster structure information. Below this threshold, dynamical mass measurements may encounter considerable difficulty in resolving the necessary information to make accurate mass predictions.

CNN$_\text{1D}$ and CNN$_\text{2D}$ have similar sampling variation curves, with  CNN$_\text{2D}$ exhibiting slightly less sensitivity to the sampling rate. $\epsilon^{(r)}$ in the CNN models show a slight correlation with sampling fraction, suggesting that the CNNs derive some information from sample richness. For a sampling rate chosen uniformly between $0.6$ and $0.8$, the sampled mass predictions for CNN$_\text{1D}$ or CNN$_\text{2D}$ can be expected to vary within a $\pm1\sigma$ interval of $85$-$102\%$ or $90$-$103\%$ of their fully sampled prediction, respectively. Both CNN models converge to negative values of $\epsilon^{(0)}$, demonstrative of a model output of $y=0$ (Equation \ref{eqn:y}). 

Figure \ref{fig:rich} infers the argument that the CNN models are less sensitive to sampling variation than either the $M$-$\sigma$ or SDM approaches.  The width of the $\epsilon^{(r)}$ scatter for the $M$-$\sigma$ models increases considerably as more cluster members are randomly removed. The $M$-$\sigma$ models do not bias away from $\epsilon^{(r)}=0$ as a result of the richness-corrected velocity dispersion estimator (Equation \ref{eqn:ustdev}). The SDM models produce higher sampling variation scatter than the $M$-$\sigma$ and also do not bias considerably from $\epsilon^{(r)}=0$.


The CNN models display the lowest six to eight sampling deviation scatter of all investigated models, reducing the six to eight residual ranges of the best $M$-$\sigma$ and SDM models by up to $30\%$. This robust behavior is primarily driven by the KDEs used to normalize the CNN model input, which are relatively insensitive to variations in sample number count. The CNN estimators presented in this paper are shown to be robust under fluctuations in the sampling rate.

\subsection{Training and Evaluation Time}

The final performance metrics we will consider are estimations of training and evaluation time. While of secondary importance to prediction error, fast implementation and execution are advantageous qualities of cluster measurements, especially when analyzing large data sets. As the abundance of high-quality data continues to increase \citep{Dodelson2016}, mass modeling methods are expected to improve computational efficiency.

In the analysis presented here, we have seen that implementation of the CNN approach is significantly faster than SDM. The full cross-validation training-and-evaluation procedure run on the catalog described in Section \ref{subsec:contam_mocks} lasts approximately 10 minutes with CNN models and 6 hr with SDM. This CNN speedup is important, especially given that the practical data sets may be orders of magnitude larger than those discussed here. 

In general, CNN models are more computationally efficient than SDMs. SDMs are nonparametric and transductive \citep{Sutherland2012}, meaning that the model complexity and evaluation procedure scale as the number of train+test points. The training and evaluation steps for SDM influence one another, implying that fitted SDM models need to be retrained upon encountering new unlabeled test data. These attributes may be undesirable in practice, where the test examples may scale up to terabytes of data. In comparison, CNN models undergo supervised, inductive learning procedures, where training and evaluation are independent calculations. The complexity of CNNs is fixed by the chosen neural architecture. In recent years, deep neural models such as CNNs have benefited from the increased use of GPUs, which speed up evaluations of neural architecture considerably \citep{LeCun2015}. CNNs find use in applications where data are overwhelmingly abundant. Under these conditions, other models such as SDM may be intractable.

\section{Conclusion} \label{sec:conclusion}
We present a novel ML method for inferring dynamical masses of galaxy clusters. Our method leverages the use of CNNs to model complex cluster substructure and to mediate systematics of traditional dynamical mass measurements. We learn cluster mass directly from distributions of galaxy kinematics, namely LOS velocity ($v_\text{los}$) and projected radial distance to the cluster center ($R_\text{proj}$). We employ KDEs to create normalized heatmap ``images" of these distributions, which serve as input to our deep neural architecture. Using this set of inputs, we train CNNs as a regression over a single output variable, the logarithmic cluster mass ($\log_{10}[M_{200c}\ (h^{-1}\mathrm{M}_\odot)]$). We then assign cluster mass predictions to unseen test data via inductive inference. This paper discusses two versions of this method, named CNN$_\text{1D}$ and CNN$_\text{2D}$ for their respective learned input spaces $\{v_\text{los}\}$ and $\{R_\text{proj},v_\text{los}\}$.

We train and evaluate our model using a catalog of realistic mock cluster observations constructed from dark matter simulation at a single redshift snapshot of $z=0.117$. The mock observations determine cluster membership via a simplistic cylindrical cut of fixed aperture ($R_\text{aperture} = 1.6\ h^{-1}\mathrm{Mpc}$) and velocity cut ($v_\text{cut} = 2200\ \mathrm{km}\ \mathrm{s}^{-1}$). We use a 10-fold cross-validation scheme to rigorously test our models on independent mock observations. We perform a comparative analysis of our models' performances with respect to several baselines including the realistic and idealized $M$-$\sigma$ and a similar ML method \citep[SDM;][]{Ntampaka2015, Ntampaka2016}. The findings of our analysis are summarized as follows:
\begin{itemize}
    \item CNN$_\text{1D}$ and CNN$_\text{2D}$ produce mass predictions with low scatter and bias in the mass range $14\leq\log_{10}[M_{200c}\ (h^{-1}\mathrm{M}_\odot)]\leq15$. We see that CNN$_\text{2D}$ reduces the error margin of CNN$_\text{1D}$ by $24\%$, suggesting that the supplemental $R_\text{proj}$ input is informative of cluster mass. Training and validation loss curves do not indicate overfitting.
    \item CNN$_\text{1D}$ and CNN$_\text{2D}$ reduce the error margin of simplistic, contaminated $M$-$\sigma$ measurements by $45\%$ and $58\%$, respectively. We compare our models to an $M$-$\sigma$ measurement with perfect member selection (pure and complete) and observe that CNN$_\text{1D}$ and CNN$_\text{2D}$ reduce prediction error by $10\%$ and $32\%$, respectively.
    \item CNN methods show improved predictive performance relative to SDM \citep{Ntampaka2015, Ntampaka2016} and other ML approaches \citep{Armitage2018}. In our comparison, CNN$_\text{2D}$ reduces the error of SDM by $17\%$.
    \item Mass predictions from CNN models have lognormal residuals. The effects of non-Gaussianity in CNN$_\text{2D}$ residuals result in a lower systematic uncertainty in cluster abundance measurements than all other investigated models ($1.7\%)$.
    \item In the context of our test catalog, the CNN models recover log-linear $M_\mathrm{pred}$-$M_\mathrm{true}$ relations with biases and scatter among the lowest measured for modern galaxy-based cluster mass estimators \citep{Wojtak2018}. However, we present this conclusion with the caveat that analyses of our models and those of other recorded methods were conducted on different data sets.
    \item CNN methods are robust under input sampling variation. Relative to $M$-$\sigma$ and SDM, predictions made by CNN models show the lowest prediction variation when inputs are randomly subsampled. This is a desirable model property, especially under conditions where some unknown fraction of galaxies are indistinguishable or otherwise not observable.
    \item For either CNN model, the 20-epoch training procedure of a single fold with $\sim$$10,000$ labeled inputs lasts about one minute. For each test input, average evaluation time can be broken down into KDE generation time ($73\ \mu$s for CNN$_\text{1D}$ and $410\ \mu$s for CNN$_\text{2D}$) and network evaluation time ($44\ \mu$s for either model). The entire training and evaluation procedure for CNN models is considerably faster than that of SDM ($\sim$$6$ hr).
\end{itemize}

We remark that the results described in this manuscript are presented in the context of the assumptions listed in Section \ref{subsec:contam_mocks}. The cluster observations used to train our model are independent from, but constructed identically to, our evaluation catalog. Our catalog construction procedure does not account for a variety of observational systematics such as obstruction, lensing, miscentering, or galaxy dark matter bias. Mass estimates produced by our model are only reliable for clusters at redshifts near that of our training catalog, as a result of redshift-dependent factors such as the definition of $M_\text{200c}$ and the distribution of interloping galaxies. Lastly, the CNN models produce singular point estimates of cluster mass and their cross-validation scatter $\sigma_\epsilon$ should not be interpreted as a Bayesian posterior. We seek to perform the above further analyses as part of a later publication.

In conclusion, mass predictions produced by CNN methods have low, lognormal error relative to other dynamical mass estimates, are stable under input sampling variation, and are computationally efficient to implement and evaluate. The CNN approach presented here may be a preferred dynamical mass estimator under conditions where high-quality simulated data is abundant or where richness measurements are uncertain or expensive. Future work involving this approach would investigate CNN modeling with more complex data inputs and deeper neural architectures. These models could potentially consolidate information from a variety of measurements (spectroscopic, X-ray, microwave, etc.) to produce a complete, precise, and unbiased prediction of cluster mass.

We thank the reviewer as well as Rachel Mandelbaum and Yizhou He for helpful input while developing this project. We thank Andrew Hearin and Peter Behroozi for preparing UniverseMachine catalogs of MDPL2 simulation data. This work is supported in part by DOE DE-SC0011114 and NSF 1563887. The computing resources necessary to complete this analysis were provided by the Pittsburgh Supercomputing Center. The CosmoSim database used in this paper is a service by the Leibniz-Institute for Astrophysics Potsdam (AIP). The MultiDark database was developed in cooperation with the Spanish MultiDark Consolider Project CSD2009-00064. 

\appendix 

\label{apx}
We describe the procedure for calculating $x_\text{proj}$, $y_\text{proj}$, $R_\text{proj}$, and $v_\text{los}$ for an arbitrary cluster-galaxy pair under the assumptions stated in Section \ref{subsec:contam_mocks}. Let  $\mathbf{r}_\text{clu}^{\text{(CM)}}$ and $\mathbf{r}_\text{gal}^{\text{(CM)}}$ represent the comoving simulation positions of the cluster and galaxy, respectively. Furthermore, define $\mathbf{r} = \mathbf{r}_\text{gal}^{\text{(CM)}} - \mathbf{r}_\text{clu}^{\text{(CM)}}$ to be the comoving distance vector between the objects. Let $\{\hat{\mathbf{x}}_\text{los}, \hat{\mathbf{y}}_\text{los}, \hat{\mathbf{z}}_\text{los}\}$ be an orthonormal basis representation of the chosen LOS, where $\hat{\mathbf{z}}_\text{los}$ is oriented along the LOS axis, and $\hat{\mathbf{x}}_\text{los}$ and $\hat{\mathbf{y}}_\text{los}$ dictate the azimuthal orientation of the observer. Under these conditions, we can write $x_\text{proj}$ and $y_\text{proj}$ for a given cluster-galaxy relationship as follows:
\begin{gather}
        x_\text{proj} = [\mathbf{r} - (\mathbf{r}\cdot \hat{\mathbf{z}}_\text{los})\hat{\mathbf{z}}_\text{los}]\cdot \hat{\mathbf{x}}_\text{los} \label{eqn:x_proj},\\
        y_\text{proj} = [\mathbf{r} - (\mathbf{r}\cdot \hat{\mathbf{z}}_\text{los})\hat{\mathbf{z}}_\text{los}]\cdot \hat{\mathbf{y}}_\text{los}.
\end{gather}
We will also often use an additional quantity called the projected radius $R_\text{proj} = (x_\text{proj}^2 + y_\text{proj}^2)^{1/2}$ which is invariant to azimuthal rotations.

To calculate $v_\text{los}$, we first find the comoving distance from the observer to the galaxy:
\begin{align}
    d^\text{(CM)}_\text{clu} &= \int_0^{z_\text{clu}} \frac{c}{H(z)}\ dz\label{eqn:d^CM}\\[2ex]
    \begin{split}
        d^\text{(CM)}_\text{gal} &= \Big| d^\text{(CM)}_\text{clu}\hat{\mathbf{z}}_\text{los} + \mathbf{r} \Big| \\
        &\approx d^\text{(CM)}_\text{clu} + \mathbf{r}\cdot \hat{\mathbf{z}}_\text{los}
    \end{split}
\end{align}
where $z_\text{clu}$ is the redshift of the cluster center. $H(z)$ is the Hubble parameter as a function of redshift and is dependent on the chosen cosmology (See Section \ref{subsec:simulation}). Since Equation (\ref{eqn:d^CM}) generally has no analytical solution, we use a numerical quadrature interpolation scheme to generate a function for $d^\text{(CM)}(z)$ and the corresponding inverse $z(d^\text{(CM)})$. The latter allows us to calculate $z_\text{gal} = z(d^\text{(CM)}_\text{gal})$ which is necessary for determining Hubble flow velocities $v^{(H)}$.
\begin{equation}
    v^{(H)}(z) = \Bigg[\frac{(1+z)^2 - 1}{(1+z)^2 + 1}\Bigg] c.
\end{equation}
Let $\mathbf{v} = \mathbf{v}^\text{(P)}_\text{gal} - \mathbf{v}_\text{clu}^\text{(P)}$ represent the relative comoving peculiar velocity between the cluster candidate and the galaxy member. We apply the small-angle approximation again to calculate peculiar velocities along the LOS $v^\text{(P,los)}$,
\begin{align}
    v^\text{(P,los)}_\text{clu} &= \mathbf{v}^\text{(P)}\cdot\Hat{\mathbf{z}}_\text{los}\\[2ex]
    \begin{split}
        v^\text{(P,los)}_\text{gal} &= \Big|\mathbf{v}_\text{clu}^{(P)} + \mathbf{v} \Big| \\
        &\approx v^\text{(P,los)}_\text{clu} + \mathbf{v}\cdot \hat{\mathbf{z}}_\text{los}.
    \end{split}
\end{align}
Equipped with these peculiar velocities and the following Hubble velocities $v^\text{(H)}_\text{clu} = v^\text{(H)}(z_\text{clu})$ and $v^\text{(H)}_\text{gal} = v^\text{(H)}(z_\text{gal})$, we can finally write an expression for $v_\text{los}$,
\begin{equation}
    v_\text{los} = \Big(v^\text{(P,los)}_\text{gal} + v^\text{(H)}_\text{gal}\Big) - \Big(v^\text{(P,los)}_\text{clu} + v^\text{(H)}_\text{clu}\Big), \label{eqn:v_los}
\end{equation}
where $\pm$ are the relativistic linear velocity addition/subtraction operators.

\end{document}